\documentclass[useAMS,usenatbib, usegraphicx]{mn2e}
\usepackage{amssymb}

\newcommand{\rmd}{{\rm{d}}}
\newcommand{\rme}{{\rm{e}}}
\newcommand{\rmi}{{\rm{i}}}

\newcommand{\vpi}{\mathbf \pi}
\newcommand{\vecd}{\mathbf d}
\newcommand{\matC}{\mathbf C}
\newcommand{\matQ}{\mathbf Q}
\newcommand{\mr}{\mathrm}

\newcommand{\cmnt}[1]{}

\newcommand{\NMC}{{459}}

\newcommand{\mytilde}{\raise.17ex\hbox{$\scriptstyle\mathtt{\sim}$}}
\newcommand{\tmin}{{\vt_{\rm min}}}
\newcommand{\tmax}{{\vt_{\rm max}}}
\newcommand{\tplog}{T_{+n}^{\rm{log}}}
\newcommand{\tmlog}{T_{-n}^{\rm{log}}}
\newcommand{\tpmlog}{T_{\pm n}^{\rm{log}}}

\newcommand{\xip}{\xi_+}
\newcommand{\xim}{\xi_-}

\newcommand{\vt}{\theta}
\newcommand{\pd}{P_\delta}

\title[Cosmic shear in SDSS]{Seeing in the dark -- II. Cosmic shear in
  the Sloan Digital Sky Survey}
\date{\today}
\author[E.~M.~Huff et al. ]{Eric~M.~Huff$^1$, Tim~Eifler$^2$, Christopher~M.~Hirata$^3$, Rachel~Mandelbaum$^{4,5}$,
  \newauthor
David~Schlegel$^6$, Uro\v s~Seljak$^{6,7,8,9}$\\
$^1$Department of Astronomy, University of California at Berkeley, Berkeley, CA 94720, USA\\
$^2$Center for Cosmology and Astro-Particle Physics, The Ohio State University, 191 W. Woodruff Avenue, Columbus, OH 43210, USA\\
$^3$Department of Astronomy, Caltech M/C 350-17, Pasadena, CA 91125, USA\\
$^4$Department of Astrophysical Sciences, Princeton University, Peyton Hall, Princeton, NJ 08544, USA\\
$^5$Department of Physics, Carnegie Mellon University, Pittsburgh,
PA 15213, USA \\
$^6$Lawrence Berkeley National Laboratory, Berkeley, CA 94720, USA\\
$^7$Space Sciences Lab, Department of Physics and Department of Astronomy, University of California, Berkeley, CA 94720, USA\\
$^8$Institute of the Early Universe, Ewha Womans University, Seoul, Korea\\
$^9$Institute for Theoretical Physics, University of Zurich, Zurich, Switzerland
}

\begin{document}
\maketitle

\begin{abstract}
  Statistical weak lensing by large-scale structure -- cosmic shear --
  is a promising cosmological tool, which has motivated the design of
  several large upcoming surveys.  Here, we present a measurement of
  cosmic shear using coadded Sloan Digital Sky Survey (SDSS) imaging
  in 168 square degrees of the equatorial region, with $r<23.5$ and
  $i<22.5$, a source number density of 2.2 per arcmin$^2$ and median
  redshift of $z_\mathrm{med}=0.52$.  These coadds were generated
  using a new method described in the companion Paper I
  \citep{2011arXiv1111.6958H} that was intended to minimise systematic
  errors in the lensing measurement due to coherent PSF anisotropies
  that are otherwise prevalent in the SDSS imaging data.  We present
  measurements of cosmic shear out to angular separations of 2
  degrees, along with systematics tests that (combined with those from
  Paper I on the catalogue generation) demonstrate that our results
  are dominated by statistical rather than systematic errors.
  Assuming a cosmological model corresponding to WMAP7
  \citep{2011ApJS..192...18K} and allowing only the amplitude of
  matter fluctuations $\sigma_8$ to vary, we find a best-fit value of
  $\sigma_8=0.636^{+0.109}_{-0.154}$ (1$\sigma$); without systematic
  errors this would be $\sigma_8=0.636^{+0.099}_{-0.137}$
  (1$\sigma$). Assuming a flat $\Lambda$CDM model, the combined
  constraints with WMAP7 are
  $\sigma_8=0.784^{+0.028}_{-0.026}$(1$\sigma$)$^{+0.055}_{-0.054}$(2$\sigma$)
  and
  $\Omega_mh^2=0.1303^{+0.0047}_{-0.0048}$(1$\sigma$)$^{+0.009}_{-0.009}$(2$\sigma$);
  the 2$\sigma$ error ranges are respectively 14\ and 17\ per cent
  smaller than WMAP7 alone.  Aside from the intrinsic value of such
  cosmological constraints from the growth of structure, we identify
  some important lessons for upcoming surveys that may face similar
  issues when combining multi-epoch data to measure cosmic shear.
\end{abstract}


\section{Introduction}

As a result of gravitational lensing, large scale inhomogeneities in
the matter density field produce small but systematic fluctuations in
the sizes, shapes, and fluxes of distant objects that are coherent
across large scales. This effect was first suggested as a tool for
constraining the form of the metric in 1966 by
\cite{1966ApJ...143..379K}. In a more modern context, the two-point
statistics of lensing fluctuations allow the only truly direct
measurement of the matter power spectrum and the growth of structure
at late times, when dark energy has caused an accelerated expansion of
the universe \citep{1998AJ....116.1009R,1999ApJ...517..565P} and
affected the growth of structure. Many studies have pointed out that
high signal-to-noise cosmic shear measurements would be
extraordinarily sensitive probes of cosmological parameters
\citep[e.g.,][]{2002PhRvD..65f3001H, 2004PhRvD..70l3515B}, which led
to its being flagged as one of the most promising probes of dark
energy by the Dark Energy Task Force
\citep{2006astro.ph..9591A}. Direct measurements of the growth of
structure also offer the opportunity to test alternative models of
gravity \citep[e.g.,][]{2011arXiv1109.4535L}.

Cosmic shear measurements were attempted as early as 1967
\citep{1967ApJ...147..864K}, but until the turn of the millennium 
\citep{2000MNRAS.318..625B,2000astro.ph..3338K,2000A&A...358...30V,2000Natur.405..143W}, no astronomical survey had the statistical power to detect it. The difficulty of the measurement is a consequence of
the near-homogeneity and isotropy of the universe. An order unity
distortion to galaxy images requires an integrated line-of-sight
matter over-density of:
\begin{equation}
\Sigma_{\rm{crit}} =  \frac{c^2}{4 \pi G} \frac{d_S}{d_L \: d_{LS}}
\end{equation}
where $d_S$, $d_L$, and $d_{LS}$ are the angular diameter distances
from the observer to the background source, from the observer to the
lens, and from the lens to the background source, respectively. A
fluctuation in the surface density $\Delta \Sigma$ leads to a shear
distortion $\gamma \sim \Delta \Sigma / \Sigma_{\rm{crit}}$.

Averaged over large ($\sim$ 100 Mpc) scales, typical line-of-sight
matter fluctuations are only $10^{-3}\Sigma_{\rm{crit}}$.  The primary
source of noise in the shear measurement, the random intrinsic
dispersion in galaxy shapes, is orders of magnitude larger; typically
the shape noise results in a dispersion in the shear of 
$\sigma_{\gamma}=0.2$. Worse, even in modern ground-based astronomical
imaging surveys, the coherent distortions -- or point-spread function
(PSF) -- induced by effects of the atmosphere, telescope optics, and
detectors are typically several times larger than the cosmological
signal (e.g., \citealt{2011arXiv1110.4913H} and Paper I in this
series). Estimating the distances to the background sources is both
crucial \citep{2006ApJ...636...21M} and difficult
\citep{2008ApJ...682...39M,2010MNRAS.401.1399B}; errors there will
modulate the amplitude of the signal through $\Sigma_{\rm{crit}}$,
biasing inference of the growth of structure.

These obstacles define the observational problem. While the existence
of cosmic shear has been established by the first studies to detect
the effect, the full potential of cosmological lensing remains to be
exploited. Few data sets capable of achieving the signal strength for
a cosmologically competitive measurement presently exist -- the
Canada-France-Hawaii Telescope Legacy Survey (CFHTLS;
\citealt{2006ApJ...647..116H,2006A&A...452...51S,2007MNRAS.381..702B,2008A&A...479....9F}),
the Cosmological Evolution Survey (COSMOS;
\citealt{2007ApJS..172..239M,2010A&A...516A..63S}), and the subset of
the SDSS imaging studied here.  However, several large surveys are
planned for the immediate and longer-term future that will
substantially expand the amount of available data for cosmological
weak lensing studies.  In the next few years, these include Hyper
Suprime-Cam (HSC, \citealt{2006SPIE.6269E...9M}), Dark Energy Survey
(DES\footnote{\texttt{https://www.darkenergysurvey.org/}},
\citealt{2005astro.ph.10346T}), the KIlo-Degree Survey
(KIDS\footnote{\texttt{http://www.astro-wise.org/projects/KIDS/}}),
and the Panoramic Survey Telescope and Rapid Response System
(Pan-STARRS\footnote{\texttt{http://pan-starrs.ifa.hawaii.edu/public/}},
\citealt{2010SPIE.7733E..12K}).  Further in the future, there are even
more ambitious programs such as the Large Synoptic Survey Telescope
(LSST\footnote{\texttt{http://www.lsst.org/lsst}},
\citealt{2009arXiv0912.0201L}),
Euclid\footnote{\texttt{http://sci.esa.int/science-e/www/area/index.cfm?fareaid\
    =102}}, and the Wide-Field Infrared Survey Telescope
(WFIRST\footnote{\texttt{http://wfirst.gsfc.nasa.gov/}}).

For this work, we have combined several methods discussed in the
literature as viable techniques for measuring cosmic shear while
removing common systematic errors. In Paper I \citep{2011arXiv1111.6958H}, we began with the
PSF model
generated by the Sloan Digital Sky Survey (SDSS) pipeline over
$\sim 250$ deg$^2$ that had been imaged many times, and 
employed a
rounding kernel method similar to that proposed in
\cite{2002AJ....123..583B}.  The result, after appropriate masking of
problematic regions, was 168 square degrees 
of deep coadded imaging with a well
controlled, homogeneous PSF and sufficient galaxy surface density to
measure a cosmic shear signal. The usable area in $r$ band was only 140
square degrees because of a PSF model error problem on the camcol 2 charge-coupled device (CCD),
which is suspected to be an amplifier non-linearity problem.

In this work, we use the catalogue from Paper I to produce a cosmic shear measurement that is dominated
by statistical errors. Section \ref{sec:model} enumerates the primary
sources of systematic error when measuring cosmic shear using our
catalogue (the properties of which are summarized briefly in
Sec.~\ref{sec:catalogues}), and describes our approaches to
constraining each of them. In Section \ref{sec:analysis_tools}, we
outline our 
 correlation function estimator and several transformations of it
that are used for systematics tests.  
Our methods for estimating covariance matrices for our observable
quantities (both due to statistical and systematic errors) are described in Sec.~\ref{sec:allcov}.  
Finally, section \ref{sec:constraints} presents the constraining power
of this measurement alone for a fiducial cosmology, and in combination
with the 7-year {\slshape Wilkinson Microwave Anisotropy Probe}
\citep[WMAP7,][]{2011ApJS..192...18K} parameter constraints to produce
a posterior probability distribution over $\Omega_m h^2$, $\Omega_b
h^2$, $\sigma_8$, $n_s$, and $w$. We show that in addition to its
value as an independent measurement of the late-time matter power
spectrum, this measurement provides some additional constraining power
over WMAP7 within the context of $\Lambda$CDM.  We conclude with some
lessons for the future in Sec.~\ref{sec:conclusions}.

While this work was underway, we learned of a parallel effort by Lin
et al. (2011). These two efforts use different methods of coaddition,
different shape measurement codes, different sets of cuts for the
selection of input images and galaxies, and analyze their final
results in different ways; what they have in common is their use of
SDSS data (not necessarily the same sets of input imaging) and their
use of the SDSS {\sc Photo} pipeline for the initial reduction of the
single epoch data and the final reduction of the coadded data
(however, they use different versions of {\sc Photo}).  Using these
different methods, both groups have extracted the cosmic shear signal
and its cosmological interpretations. We have coordinated submission
with them but have not consulted their results prior to this, so these
two analysis efforts are independent, representing versions of two
independent pipelines.

\section{Catalogues}\label{sec:catalogues}

Paper I \citep{2011arXiv1111.6958H} describes a coadd imaging dataset,
optimised for cosmic shear measurement, constructed from single-epoch
SDSS images in the Stripe 82 equatorial region, with right ascension
(RA) $-50^\circ<$RA$<+45^\circ$, and declination
$-1.25^\circ<$Dec$<+1.25^\circ$. In that work, we apply an adaptive
rounding kernel to the single-epoch images to null the effects of PSF
anisotropy and match to a single homogeneous PSF model for the entire
region, and show that in the resulting shear catalogues, the amplitude
of the galaxy shape correlations due to PSF anisotropy at angular
separations greater than 1 arcminute is negligible compared to the
expected cosmic shear statistical errors.

The final shape catalogue described in that work consists of 1\,067\,031
$r$-band and 1\,251\,285 $i$-band shape measurements with characteristic
limiting magnitudes of $r<23.5$ and $i<22.5$, over effective areas of
140 and 168 square degrees, respectively.

\section{Model for the lensing and systematic error signals}
\label{sec:model}

We model the observed galaxy shape field as the sum of a cosmic shear
component, an independent systematics field produced by anisotropies
in the effective PSF $e_\mathrm{psf}$, and a systematics field
produced by the intrinsic spatial correlations of galaxy shapes
$e_\mathrm{int}$ (intrinsic alignments; e.g.,
\citealt{2004PhRvD..70f3526H}). We allow for a shear calibration
factor that depends on the shear responsivity $\mathcal{R}$
\citep{2002AJ....123..583B} of the ensemble of galaxy surface
brightness profiles to the underlying gravitationally-induced shear
$\gamma$. Typically $\mathcal{R}\approx 1-e_\mathrm{rms}^2$, however
we consider it to be a more general factor that also includes any
biases due to effects such as uncorrected PSF dilution, noise-related
biases, or selection biases. We assume that the galaxy shape
response to PSF anisotropies $\mathcal{R}_{{\rm psf}}$ is not {\em a
  priori} known, but rather suffers from a similar set of
`calibration' uncertainties as the response of the ensemble of galaxy
images to gravitational lensing shear.  Thus we define our model for
the two ellipticity components ${\bmath e} = (e_1, e_2)$ as
\begin{equation} \label{eq:emodel}
  {\bmath e} = \mathcal{R} {\bf \gamma} +
  \mathcal{R}_{\rm psf}{\bmath e_{{\rm psf}}} + {\bmath e_{\rm int}}.
\end{equation}

We assume that the two-point statistics of the underlying (cosmological) shear field
$\langle\gamma \gamma\rangle$ consist entirely of $E$-modes, $e_{\gamma,E}$ (which is a good enough
approximation given the size of our errors; \citealt{2002ApJ...568...20C,2002A&A...389..729S}), and are
statistically independent of the PSF when averaged over large
regions. We also assume that the PSF and the
intrinsic alignments are independent -- but not that the lensing shear
and intrinsic alignments are independent \citep{2004PhRvD..70f3526H}. The two-point correlation
of the galaxy shapes contains terms resulting from gravitational
lensing and from
systematic errors:
\begin{equation}
  \langle{\bmath ee}\rangle = 4\mathcal{R}^2 \xi_{\gamma,E} +
  \mathcal{R}_\mathrm{psf}^2 \xi_\mathrm{psf}+\xi_\mathrm{int}+\langle
  \bgamma{\bmath e}_\mathrm{int}\rangle.
\label{eq:model}
\end{equation}
Here, $\xi_\mathrm{psf}$ is the auto-correlation of the PSF ellipticity
field.  Errors in the determination of the galaxy redshift
distribution will enter as a bias in the predicted $\xi_{\gamma,E}$.

Our goal is to carry out a statistics-limited measurement of
$\xi_{\gamma,E}$. This will entail showing that the combined
amplitudes of $\mathcal{R}_\mathrm{psf}^2 \xi_\mathrm{psf}$,
$\xi_\mathrm{int}$, $\langle\bgamma \,{\bmath e}_\mathrm{int}\rangle$,
the uncertainty in the theoretically-predicted $\xi_{\gamma,E}$
arising from redshift errors, and the uncertainty in the shear
calibration (via the responsivity $\mathcal{R}$) contribute less than
$20$ per cent to the statistical errors in $\langle{\bmath
  ee}\rangle$.

Our approach to handling of systematic error is as follows: we attempt
to reduce each systematic to a term that can be robustly and
believably estimated from real data (either the data here or in other,
related work), and we then explicitly correct for it.  These
corrections naturally have some uncertainty associated with them,
which we use to derive a systematic error component to the covariance
matrix.  The exception to the rule given here is if there is a
systematic error for which there is no clear path to estimating its
magnitude, then we do not attempt any correction, and simply
marginalize over it by include an associated uncertainty in the
covariance matrix.

\subsection{Cosmic shear}

Foreground anisotropies in the matter distribution along the line of
sight to a galaxy will generically distort the galaxy image. For 
weak lensing, the leading order lensing contribution to galaxy shapes
can be thought of as arising from a linear transformation of the image
coordinates ${\mathbfss A}{\bmath x}_{\rm true} =  {\bmath x}_{\rm obs}$, where 
\begin{equation}
{\mathbfss A}=\left(
\begin{array}{cc}
  1+\kappa + \gamma_1 & \gamma_2 \\
  \gamma_2  & 1 + \kappa - \gamma_1
\end{array}
\right).
\end{equation}

The convergence $\kappa$ causes magnification, whereas the shear
components $\gamma_1$ and $\gamma_2$ map circles to ellipses.  The
shear is related to the projected line-of-sight matter distribution,
weighted by the lensing efficiency:
\begin{equation}\label{eq:sheardef}
  \left(\gamma_1,\gamma_2\right)= \partial^{-2} \int_0^{\infty} W\left(\chi,\chi_i\right) \left(\partial_x^2-\partial_y^2,2\partial_x\partial_y\right)\delta\left(\chi\hat{\bf n}_i\right){\rm d}\chi.
\end{equation}
Here we integrate along the comoving line-of-sight distance $\chi$
(where $\chi_i$ is the distance to the source), and the matter
over-density $\delta = (\rho-\overline{\rho})/\rho$. The window
function in a flat universe is
\begin{equation}
W(\chi,\chi_i) = \frac32\Omega_{m}H_0^2(1+z)\chi^2 \left( \frac1\chi - \frac1{\chi_i} \right).
\end{equation}

The two-point correlation function of the shear can be calculated by
identifying pairs of source galaxies, and defining shear components
$(\gamma_t,\gamma_x)$ for each one to be the shear in the coordinate
system defined by the vector connecting them, and in the $\pi/4$
rotated system.  This two-point correlation function can be expressed
as a linear transformation of the matter power spectrum $P_{\delta}$
averaged over the line of sight to the sheared galaxies:
\begin{eqnarray}
\xi_{\pm} &=&\left\langle \gamma_t\gamma_t\right\rangle \pm \left\langle\gamma_{\times}\gamma_{\times}\right\rangle \nonumber \\
  &=& \frac{1}{2\pi}\int_0^{\infty}\rmd\ell \,\ell\, P_{\kappa}\left(\ell\right)J_{0,4}\left(\ell \theta\right) \label{eq:xigamma}
\end{eqnarray}
and
\begin{eqnarray}
  P_{\kappa} &=
  &\left(\frac{3\Omega_m}{2d_H^2}\right)\int_0^{\infty}\frac{\rmd\chi}{a\left(\chi\right)^2}
  P_{\delta}\left(\frac{\ell}{d\left(\chi\right)}\right) 
  \nonumber \\ 
&& \times \left[\int_{\chi}^{\infty}{\rm d}\chi'n\left(\chi'\right)\frac{d\left(\chi'-\chi\right)}{d\left(\chi'\right)}\right]^2,
\label{eq:pkappa}
\label{eqn:shear_xi}
\end{eqnarray}
where the last expression makes use of Limber's approximation and
$d(\chi)$ is the distance function, i.e. $\chi$ in a flat universe,
$K^{-1/2}\sin K^{1/2}\chi$ in a closed universe, and $(-K)^{-1/2}\sinh
(-K)^{1/2}\chi$ in an open universe. In the expression in brackets,
$n(\chi')$ represents the source distribution as a function of
line-of-sight distance (normalised to integrate to 1).
This statistic ($P_\kappa$) is sensitive both to the distribution of
matter $\delta$ and to the background cosmology, via both the explicit
$\Omega_m$ dependence and the distance-redshift relations.

\subsection{Intrinsic alignments}
\label{subsec:ia}

Many studies have discussed intrinsic alignments of galaxy shapes due
to effects such as angular momentum alignments or tidal torque due to
the large-scale density field \cite[for pioneering studies,
see][]{2000ApJ...545..561C,2000MNRAS.319..649H,2001MNRAS.320L...7C,2001ApJ...559..552C,2002MNRAS.335L..89J,2005ApJ...618....1H}. While
these effects can generate coherent intrinsic alignment 2-point
functions, \cite{2004PhRvD..70f3526H} pointed out that the large-scale
tidal fields that can cause intrinsic alignments are sourced by the
same large-scale structure that is responsible for producing a cosmic
shear signal. Thus, in this model, the intrinsic alignments do not
just have a nonzero auto-correlation, they also have a significant
anti-correlation with the lensing shear which can persist to very
large transverse scales and line-of-sight separations. If left
uncorrected, this coherent alignment of intrinsic galaxy shapes
suppresses the lensing signal, since the response of the intrinsic
shape to an applied tidal field has the opposite sign from the
response of the galaxy image to a shear with the same magnitude and
direction. We generally refer to the intrinsic alignment
auto-correlation as the ``$II$'' contamination and its correlation
with gravitational lensing as the ``$GI$'' contamination. This can be
compared to the pure gravitational lensing auto-correlation
(``$GG$'').

To address the uncertainty related to intrinsic alignments, 
we rely on empirical measurements that constrain the degree to which
they might affect our measurement.  Several studies using SDSS imaging
and spectroscopic data
\citep[e.g.,][]{2006MNRAS.367..611M,2007MNRAS.381.1197H,2009ApJ...694..214O,2011A&A...527A..26J,2011MNRAS.410..844M}
have demonstrated the existence of intrinsic alignments of galaxy
shapes on cosmological distance scales.  \cite{2007MNRAS.381.1197H}
used the luminosity and colour-dependence of intrinsic alignments for
several SDSS galaxy samples to estimate the contamination of the
cosmic shear signal due intrinsic alignments for lensing surveys as a
function of their depth.  These estimates were a function of the
assumptions that were made, for example about evolution with redshift.
The ``central'' model given in that paper leads to a fractional
contamination of
\begin{equation}
\frac{ C_{\ell=500, GI} }{ C_{\ell=500, GG} } \approx -0.08
\label{eqn:gi_signal}
\end{equation}
for a limiting magnitude of $m_{R,\mathrm{lim}}=23.5$, which is close
to the limiting magnitude of our sample.  Subsequent work
\citep{2011A&A...527A..26J,2011MNRAS.410..844M} provided more
information about redshift evolution; primarily those results were in
broad agreement with the previous ones, and were sufficient to rule
out both the ``optimistic'' and the ``very pessimistic'' models in
\cite{2007MNRAS.381.1197H}.

We thus adopt the ``central'' model, and apply the correction given in
Eq.~(\ref{eqn:gi_signal}) to our theory predictions for the $C_{\ell}$
due to cosmic shear, multiplying the predicted cosmic shear power
spectrum by $0.92$ before transforming into the statistics that are
used for the actual cosmological constraints\footnote{While the
  intrinsic alignment contamination is in principle scale-dependent,
  the plots in \cite{2007MNRAS.381.1197H} suggest that this scale
  dependence is in fact quite weak for the scales used for our
  analysis, so we ignore it here.}.  We also assume this correction
has a conservative systematic uncertainty of 50 per cent, which
amounts to an overall 4 per cent uncertainty in the theory prediction
(see Sec.~\ref{sec:allcov} for a quantitative description of how we
incorporate this and other systematic uncertainties into the
covariance matrix).

Since the $GI$ correlation is first order in the intrinsic alignment
amplitude, while the $II$ power is second order, we expect the first
to be the dominant systematic. In principle, the $GI$ effect could be
smaller than $II$ if the correct alignment model is quadratic in the
tidal field rather than linear \citep{2004PhRvD..70f3526H}. However,
in the aforementioned cases in which intrinsic alignment signals are
detected at high significance (i.e. for bright ellipticals) the linear
model for intrinsic alignments appears to be valid
\citep{2011JCAP...05..010B}. Therefore we attempt no correction for
$II$.

\subsection{Shear calibration}
\label{subsec:shearcalib}

Another source of systematic error for weak lensing measurements is
uncertainty in the {\em shear calibration} factor. The galaxy
ellipticity $(e_+,e_\times)$ observed after isotropizing the PSF need
not have unit response to shear: in general, averaged over a
population of sheared galaxies, we should have
\begin{equation}
\left\langle (e_+,e_\times) \right\rangle = \mathcal{R}(\gamma_+,\gamma_\times),
\end{equation}
where $\mathcal{R}$ is the shear responsivity. It depends on both the
shape measurement method {\em and} the galaxy population
\citep[e.g.][]{2007MNRAS.380..229M, 2010MNRAS.406.2793B,
  2011MNRAS.414.1047Z}.

For this work, we used the re-Gaussianization method
\citep{2003MNRAS.343..459H}, which is based on second moments from
fits to elliptical Gaussians, and has been previously applied to SDSS
single-epoch imaging \citep{2005MNRAS.361.1287M,
  2011arXiv1110.4107R}. For this class of methods, in the absence of
selection biases and weighting of the galaxies, perfectly homologous
isophotes, and no noise, there is an analytic expectation
\citep{2002AJ....123..583B}:
\begin{equation}
\mathcal{R} = 2( 1 - e_{\rm rms}^2),
\label{eq:perfectcal}
\end{equation}
where $e_{\rm rms}$ is the root-mean-square ellipticity per component ($+$ or $\times$).

The calibration errors for re-Gaussianization and other
adaptive-weighting methods are well-studied in the literature
\citep[e.g.,][]{2004MNRAS.353..529H,2005MNRAS.361.1287M,2011arXiv1107.4629M,2011arXiv1110.4107R}. They
arise from all of the deviations from the assumptions of
Eq.~(\ref{eq:perfectcal}). Higher-order departures from
non-Gaussianity in the galaxy light profile cause errors in the PSF
dilution correction. Errors in the measurement of the PSF model will
cause a similar error in the dilution correction. The resolution
factor of an individual galaxy depends on its ellipticity, so any
resolution cut on the galaxy sample will introduce a shear bias in the
galaxy selection function.  Due to the non-linearity of the shear
inference procedure, noise in the galaxy images causes a bias in the
shears (rather than just making them noisier).  The estimation of the
shear responsivity, or even of $e_{\rm rms}$, is another potential
source of error, as the response of the galaxies to the shear depends
on the true, intrinsic shapes, rather than the gravitationally
sheared, smeared (by the PSF), noisy ones that we observe.

Past approaches to this problem have included detailed accounting for
these effects one by one. In this paper, we instead use detailed
simulations of the image processing and shape measurement pipelines,
including real galaxy images, to estimate both the shear calibration
and the redshift distribution of our catalogue. The advantage is that
this includes all of the above effects and avoids uncertainties
associated with analytic estimates of errors. The {\sc Shera} (SHEar
Reconvolution Analysis) simulation
package\footnote{\texttt{http://www.astro.princeton.edu/\mytilde{}rmandelb/shera/shera.html}}
has been previously described \citep{2011arXiv1107.4629M} and applied
to single-epoch SDSS data for galaxy-galaxy lensing
\citep{2011arXiv1110.4107R}, but this is its first application to
cosmic shear data.

To simulate our images, we require a fair, flux-limited sample of any
galaxies that could plausibly be resolved in our coadd imaging,
including high-resolution images with realistic
morphologies\footnote{Simple models with analytic radial profiles and
  elliptical isophotes are not adequate to measure all sources of
  systematic error such as under-fitting biases or those due to
  non-elliptical isophotes \citep{2010MNRAS.406.2793B}.}.  For this
purpose we use a sample of 56\,662 galaxy images drawn from the
COSMOlogical {\sc e}volution Survey
\citep[COSMOS:][]{2007ApJS..172..196K,2007ApJS..172....1S,2007ApJS..172...38S}
imaging catalogues. The deep {\slshape Hubble Space Telescope} (HST)
Advanced Camera for Surveys/Wide Field Camera (ACS/WFC) imaging in
$F814W$ (``broad $I$'') in this 1.6 deg$^2$ field is an ideal source
of a fairly-selected galaxy sample with high resolution, deep
images\footnote{Admittedly there may be some sampling variance that
  affects the morphological galaxy mix.}.  These images consist of two
samples -- a ``bright'' sample of 26\,116 galaxies in the magnitude
range $I<22.5$, and a ``faint'' sample consisting of the $22.5<I<23.5$
galaxies. The charge transfer inefficiency-corrected
\citep{2010MNRAS.401..371M} and multi-drizzled \citep[][to a pixel
scale of 0.03\arcsec]{2002hstc.conf..337K,2007ApJS..172..203R} galaxy
postage stamp images have been selected to avoid CCD edges and
diffraction spikes from bright stars, and have been cleaned of any
other nearby galaxies, so they contain only single galaxy images
without image defects.  The bright sample is used for ground-based
image simulations in \cite{2011arXiv1107.4629M}; the faint sample is
selected and processed in an identical way\footnote{We thank Alexie
  Leauthaud for kindly providing these processed images.}.  Each postage stamp is
assigned a weight to account for the relative likelihoods of
generating postage stamps passing all cuts (avoidance of CCD edges and
bright stars) for galaxies of different sizes in the COSMOS field;
this weight is calculated empirically, by comparing the size
distribution of galaxies with postage stamps to the size distribution
of a purely flux-limited sample of galaxies.

Each of these postage-stamp images has several properties associated
with it that are of interest for this analysis. The COSMOS photometric
catalogues \citep{2009ApJ...690.1236I} contain HST $F814W$ magnitudes as
well as photometric redshifts and Subaru $r-i$ colours based on
PSF-matched aperture magnitudes.

In order to simulate our observations, we first select a coadd `run'
consisting of five adjacent frames in the scan direction at random
from the list of completed runs. We draw 1250\ galaxies (exactly 250
per frame) at random from the list of COSMOS postage stamps according
to the weights described above, up-weighting the probability of
drawing the faint galaxies by a factor of 1.106\ to account for the
fact that we have sampled the faint population at a lower rate than
the bright one in constructing the image sample.

Once a list of postage-stamp images is selected, we assign $r$- and
$i$-band magnitudes by re-scaling each image; each galaxy image is
inserted into the coadded imaging with the flux it would have been
observed to have in SDSS before the addition of pixel noise. The $i$-band is chosen to be $0.03$
magnitudes brighter than the COSMOS $F814W$ ($I$) band {\tt MAG\_AUTO}
values; this small offset is based on empirical comparison with SDSS
magnitudes for brighter galaxies, to account for slight differences in
the $F814W$ and $i$ passbands \citep{2011arXiv1107.4629M}.  The
$r$-band is chosen so as to match the Subaru PSF-matched aperture
colours for each object. Each postage stamp is assigned a random,
uniformly-sampled position in the coadd run, with the postage stamps
distributed equally among the frames.

We use the {\sc shera} code to pseudo-deconvolve the HST point-spread
function, apply (if necessary; see below) a shear to each galaxy,
reconvolve each image with the known coadd point-spread function,
renormalise the flux appropriately, and resample from the COSMOS pixel
scale to the coadd pixel scale before adding that postage stamp to the
coadd image. This procedure, suggested by \cite{2000ApJ...537..555K}
and implemented to high precision in \cite{2011arXiv1107.4629M}, can
be used to simulate ground-based images with a shear appropriately
applied, despite the space-based PSF in the original COSMOS images,
and with a user-defined PSF.

The normal coadd masking algorithm is then applied, and shear
catalogues are generated as in Paper I by running the SDSS object
detection and measurement pipeline, 
{\sc Photo-frames}, followed by the shape measurement code described
in Sec.~\ref{subsec:shearcalib}. The output catalogues
are matched against the known input object positions, and a simulation
catalogue of the matches is created. We employ these simulations below
to determine the shear calibration and as an independent validation of
our inferred redshift distribution.

\begin{figure}
\includegraphics[angle=-90,width=3.2in]{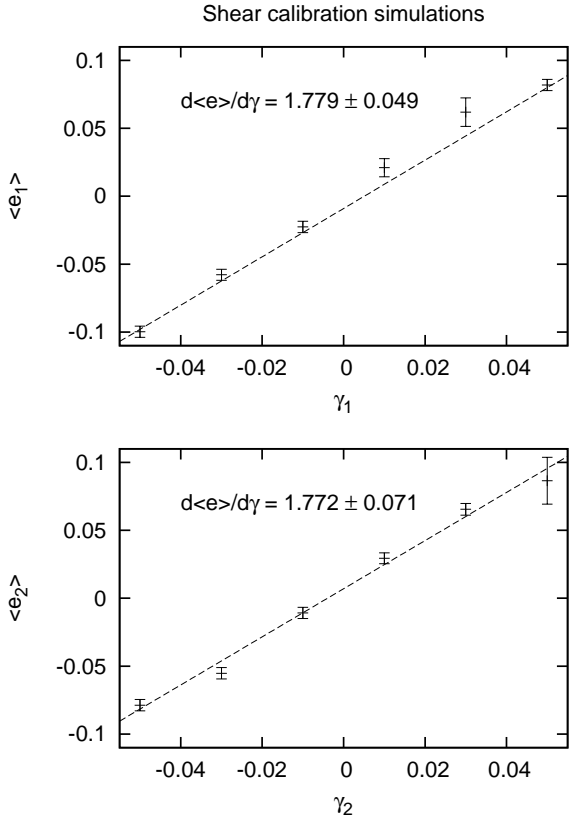}
\caption{\label{fig:shearcalib}The response of the mean ellipticities
  $\langle e_1\rangle$ and $\langle e_2\rangle$ to applied shear, as
  determined in the {\sc shera}-based simulations. Poisson error bars
  are shown. The additive offset to the response curve is not shown in
  the fit; these simulations do not accurately measure an additive
  shear bias.}
\end{figure}

For each suite of simulation realisations, we use
the same random seed (i.e., we select the same galaxies from our
catalogue and place them at identical locations in the coadded image)
but with different applied shears per component ranging from $-0.05$ to $+0.05$.
We measure the mean weighted shape of the
detected simulation galaxies produced by our pipeline, and fit a line
to the results. 
Since the same galaxies are used without
rotation, only the slope and not the intercept is meaningful. The shear
response in each component for each applied shear is shown in
Fig.~\ref{fig:shearcalib}. The responsivities in the two components are
consistent, which is expected on oversampled data with a rounded PSF.
(The unequal size of the error bars reflects the number of runs that we were
able to process by the time the shear calibration solution was frozen.)
The total number of galaxies in the final simulated catalogues was 130\,063.
The response appears to be linear for small
applied shears. Based on these results, we adopt a shear responsivity
for this galaxy population of $1.776\pm 0.043$.

\subsection{Redshift distribution}

The explicit dependence of the shear signal in
Eqs.~(\ref{eq:sheardef}) and~(\ref{eq:pkappa}) on the
distribution of lensed galaxy redshifts, 
combined with the practical impossibility of acquiring a spectroscopic
redshift for the millions of faint galaxies statistically necessary
for a cosmic shear measurement, can be a troublesome source of bias
and systematic uncertainty for cosmic shear measurements.

An error in the estimated redshift distribution leads to an
incorrect prediction for the amplitude of the shear signal at a given
cosmology. This is similar in principle to the bias arising in the
amplitude of the galaxy-galaxy lensing signal due to photometric
redshift biases explored in \cite{2011arXiv1107.1395N}; uncorrected,
standard photometric redshift estimation techniques can lead to biases in the predicted lensing signal at the $\sim
10$ per cent level.  For cosmic shear
measurements, an imperfect estimate of the redshift distribution leads
to biases in $\sigma_8$ and $\Omega_m$ that are comparable in
amplitude to the errors in the estimated mean of the redshift
distribution \citep{2006APh....26...91V}.

As a fiducial reference, the redshift distribution of the single-epoch
SDSS imaging catalogue is established to approximately 1 per cent
\citep{2011arXiv1109.5192S}; for deeper surveys over a smaller area,
this becomes a more difficult problem, as the spectroscopic
calibration samples available for inferring the redshift distribution
are limited in their redshift coverage and widely dispersed across the
sky. We employ a colour-matching technique similar to that employed by
\cite{2011arXiv1109.5192S}; in what follows, we describe the
technique, our estimate of its uncertainty, and several cross-checks
on the results.

\subsubsection{Fiducial redshift distribution}
\label{sss:zdist}

The source redshift distribution used in our analysis is derived
following \cite{2008MNRAS.390..118L} and \cite{2009MNRAS.396.2379C},
and is similar in spirit to \cite{2011arXiv1109.5192S}; the principle
is that, for two galaxy samples that span broadly similar ranges in
redshift, colour, and limiting magnitude, matched colour samples
correspond to matched redshift distributions.

Our spectroscopic calibration sample is composed of 12\,360\ galaxies,
from the union of the VIMOS VLT Deep Survey
\citep[VVDS]{lefevre_etal:2005} 22h field, the DEEP2 Galaxy Redshift
Survey \citep{davis_etal:2003, madgwick_etal:2003}, and portions of 
the PRism MUlti-object Survey (PRIMUS; \citealt{2011ApJ...741....8C},
Cool et~al. 2011 {\em in prep.}).  We follow the procedures outlined
in \cite{2011arXiv1107.1395N} for selecting good quality spectroscopic
redshifts, and avoiding duplicate galaxies in samples that overlap
(such as DEEP2 and PRIMUS). Each of
these samples has a redshift distribution that is likely to differ substantially from the redshift
distribution of our lensing catalogue: the DEEP2 catalogue in the
fields we use at $23^h30^m$ and $02^h30^m$ is heavily
colour-selected (in non-SDSS bands) towards objects at $z>0.7$; 
the PRIMUS catalogue includes several fields,
some of which are selected from imaging with a shallower limiting magnitude;
and the VVDS catalogue is selected in the $I$ band ($I<22.5$) with a
relatively high redshift
failure rate that exhibits some colour dependence. 

We assign a redshift from a galaxy in the union calibration sample to
the closest galaxy in the lensing catalogue within 3 arcsec, finding
12$\,$360 matches.  To generate a representative training sample of
galaxies from the lens catalogue, we draw $4\times 10^5$ galaxies with
replacement from the full area (not just in these regions), with
sampling probability proportional to the mean of the weights assigned
in the $r$ and $i$ bands to that galaxy for the correlation analysis
(Eq.~\ref{eqn:weight}).  Note that this procedure does not incorporate
those galaxies in the excluded camcol 2 region.

\begin{figure}
\includegraphics[width=7cm]{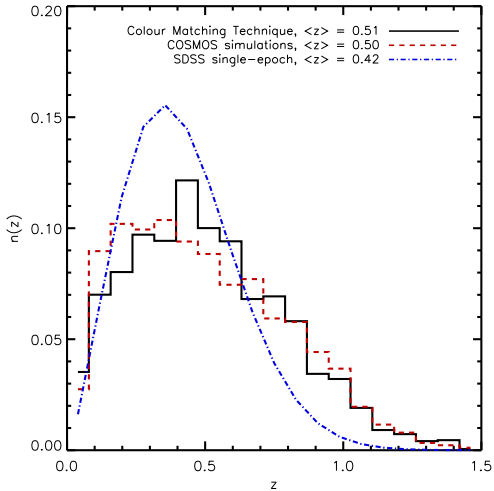}
\caption{\label{fig:nz} The redshift distribution inferred from
  matching the colours of the spectroscopic calibration sample to
  those of the lensing catalogue (solid black line,
  Sec.~\ref{sss:zdist}) shown alongside the noisier redshift
  distribution inferred from the shear calibration simulations (dashed
  red line, Sec.~\ref{subsubsec:othertests}). The best-fit
  distribution for the single-epoch SDSS lensing catalogue from
  \protect\cite{2011arXiv1107.1395N} is shown for reference as the
  blue dot-dashed line.}
\end{figure}

We use the \citet{2008MNRAS.390..118L}
code\footnote{\texttt{http://kobayashi.physics.lsa.umich.edu/\mytilde{}ccunha/nearest/}}
to solve for a set of weights over the calibration sample, such that
the re-weighted 5-dimensional magnitude distributions of the
calibration sample match those of the representative random subset of
the lensing catalogue. The histogram of the calibration sample
redshifts reweighted in this manner is shown as a solid line in
Fig.~\ref{fig:nz}. The inferred mean redshift is 0.51; in contrast to
the redshift distribution for single-epoch imaging, there is a
non-negligible fraction of the galaxy sample above $z>0.7$. We use the
solid curve based on the colour-matching techniques to calculate the
shear covariance matrix, and to predict the shear correlation function
for any given cosmology.

\subsubsection{Uncertainty}
\label{sss:z-uncert}

We expect that the primary source of error in the redshift
distribution as estimated from the combined calibration sample is
sample variance, resulting from the finite volume of the calibration
sample. To estimate its magnitude, we use the public code of
\citet{2011ApJ...731..113M} for estimating the cosmic variance of
number counts in small fields.

Our redshift binning scheme has 19 bins between $0 < z < 1.5$.  For a
collection of disparate calibration fields, we use the
\citet{2011ApJ...731..113M} code to produce a fractional error in the
number counts $\sigma_{\mathrm{gg},i,j}$ for the $j$th redshift bin in
the $i$ field (where fields are distinguished by their coverage area)
in bins of stellar mass.

The redshift sampling rate of each distinct survey in the calibration
sample differs, and so the balance of contributions to the final
redshift distribution will change as well. To account for this, we sum
over every calibration field's contribution to the reweighted redshift
distribution in the $j$ bin to estimate an absolute (not relative) overall error:
\begin{equation}
\sigma_{j}^2 =  \sum_i \left(\sigma_{\mathrm{gg},i,j} n_{\mathrm{eff},i,j}\right)^2
\end{equation}
where the effective number of galaxies contributed in the $j$ bin by
the $i$ survey is just the sum over the nearest-neighbour derived
weights assigned to calibration sample galaxies $k$ in that field $i$
and bin $j$:
\begin{equation}
n_{\mathrm{eff},i,j} = \sum_k w_{nn,i,j,k}
\end{equation}

To propagate these errors into the covariance matrix for $\xi_{E}$, we
first fit a smooth function of the form 
\begin{equation}
n_{z}\left(z\right) \propto z^a \mathrm{e}^{-\left(z/z_0\right)^b}
\end{equation}
to the nearest neighbour weighting-derived redshift distribution shown
in Figure \ref{fig:nz}; the best fit parameters are $a=0.5548$,
$z_0=0.7456$, and $b = 2.5374$. We perturb this smooth distribution by
adding a random number drawn from a normal distribution with mean
$n_z\left(z_j\right)$ (normalised to the weighted number of
calibration galaxies in that bin) and standard deviation
$\sigma_j$ 
at the location of the $j$th redshift bin.  We then renormalise the
perturbed distribution to unity, and compute the predicted cosmic
shear signal. The covariance matrix of 402 realisations of this
procedure is added to the statistical covariance matrix.

\subsubsection{Other tests}\label{subsubsec:othertests}

As an independent check on the redshift distribution, we also use the
shear calibration simulations (Sec.~\ref{subsec:shearcalib}) to
constrain the redshift distribution of our sources. The COSMOS
photometric redshifts, inferred as they are from many more imaging
bands (typically with deeper imaging) than for the SDSS data discussed
here, are very accurate.  For example, \cite{2009ApJ...690.1236I}
finds a photo-$z$ scatter of $\sigma_z/(1+z)\sim 0.01$ for a galaxy
sample with the flux limit of the SDSS coadds.  In contrast,
\cite{2011arXiv1107.1395N} found that in the SDSS single-epoch
imaging, the scatter defined in the same way was $\sim 0.1$ despite
the brighter flux limit of the single-epoch imaging (due in part to
the more limited number of bands, but primarily to the far lower
signal-to-noise ratio). If we treat the COSMOS photometric redshifts
as we would spectroscopic data, then the redshift distribution of
COSMOS galaxies that pass successfully into the shear catalogue is the
same as that of our source catalogue -- assuming, of course, that the
COSMOS field is representative of the whole of Stripe 82. It is not,
of course; large-scale structure in the COSMOS field \citep[which can
be significant, as COSMOS covers only 1.7 square
degrees;][]{2010ApJ...708..505K} can bias a determination of the
redshift distribution in this manner.  The $n(z)$ inferred from the
COSMOS-based simulations is also shown in Fig.~\ref{fig:nz}, and
agrees extremely well with the fiducial $n(z)$ derived from
colour-matching.

A final (but obviously not independent) sanity check is to compare to
the COSMOS Mock Catalogue \citep{2009A&A...504..359J}, which is being
used extensively to plan future dark energy programmes, using the cuts
$r_{\rm eff}>0.47$ arcsec, limiting magnitudes $r<23.5$, and $i<22.5$
(see Paper I, where we argue that these most closely mimic the cuts in
our data). This test predicts $\langle z\rangle = 0.51$, identical to
that obtained via the re-weighting procedure. Given the crudeness of
the procedure for comparing the results, this is an excellent
validation of the COSMOS Mock Catalogue as a forecasting tool.

\subsection{Stellar contamination}
\label{ss:stellar}

Stellar contamination of the galaxy catalogue reduces the apparent
shear by diluting the signal with round objects that are not sheared
by gravitational lensing. Because the image simulations described in
Sec.~\ref{subsec:shearcalib} only included galaxies, the resulting shear responsivities
do not include signal dilution due to accidental inclusion of stars in
the galaxy sample.  In Paper I, we estimated the stellar contamination
by comparison with the DEEP2 target selection photometry (which is
deeper and was acquired at the Canada-France-Hawaii Telescope under much better seeing conditions than typical for SDSS), and found a contamination fraction of 0.017. We also argued that the mean stellar density in the stripe must be larger than in the high-latitude DEEP2 fields, by a factor as large as 2.8. We therefore conservatively take the stellar contamination fraction $f_{\rm star}$ to be
\begin{equation}
f_{\rm star} = 0.017 (1.9\pm 0.9) = 0.032\pm 0.015.
\end{equation}
The resulting suppression of the cosmic shear signal is treated in much the same way as for intrinsic alignments: we reduce the theory signal by a factor of $(1-0.032)^2=0.936$, and add a contribution to the covariance of 0.030 times the theory signal.

\subsection{Additive systematics}
\label{subsec:additive}

Among the most worrying systematics in the early detections of cosmic
shear was additive power. This comes from any non-cosmological source
of fluctuations in shapes such as PSF anisotropy that add to the
ellipticity correlation function of the galaxies. Such power was
clearly detected in Paper I in the form of systematic variation of
both star and galaxy $e_1$ as a function of declination. The sense of
the effect -- a negative contribution to $e_1$ (in $r$ band we
have\footnote{The $1\sigma$ Poisson uncertainty in these numbers is
  0.0005 (0.0004) per component in $r$ ($i$) band.} $\langle
e_1\rangle = -0.0018$ and $\langle e_2\rangle = +0.0004$, while in $i$
band $\langle e_1\rangle = -0.0022$ and $\langle e_2\rangle =
-0.0002$) -- is suggestive of {\em masking bias}, in which the
selection of a galaxy depends on its orientation, with galaxies
aligned in the along-scan direction ($e_1<0$) being favoured, and with
no effect on $e_2$ (consistent with zero mean over the whole survey).
The reason for this particular sign is seen in Figure 2 of Paper I; as
shown, bad columns along the scan direction tend to be repeated at the
same location in multiple images, resulting in significant
(non-isotropic) masks with that directionality. Direct evidence for
masking bias comes from the change in mean ellipticity due to
increased masking: when we removed from the coadded image pixels that
were observed in fewer than 7 input runs and reran {\sc Photo-Frames},
the $\langle e_1\rangle$ signal became {\em worse}: $-0.0051$ in $r$
band and $-0.0044$ in $i$ band, whereas $\langle e_2\rangle$ was
essentially unchanged. This increase is difficult to explain in terms
of spurious PSF effects, so we conclude that our galaxy catalogue
likely contains a mixture of masking bias as well as possible additive
systematics from PSF ellipticity in the coadded image.

\begin{figure}
\includegraphics[angle=-90,width=3.2in]{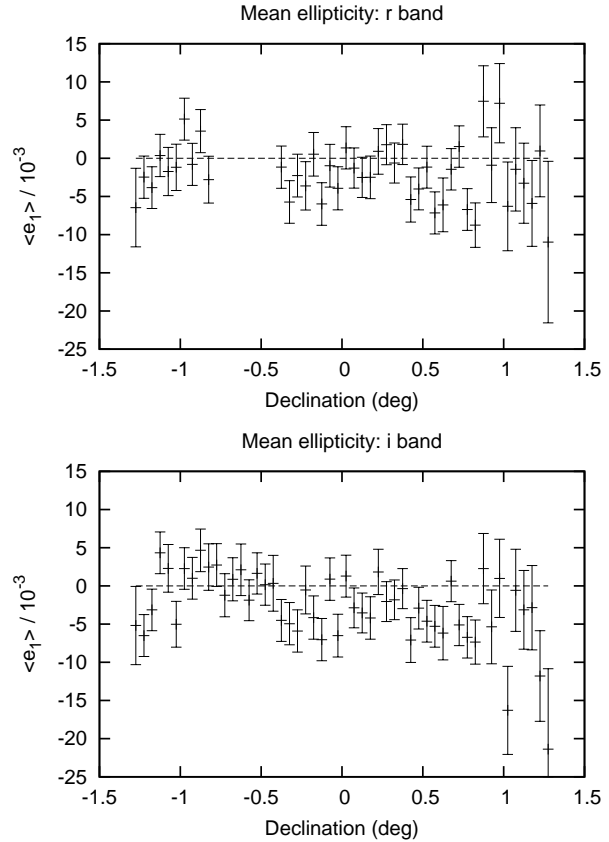}
\caption{\label{fig:mean_e1_plot} The mean ellipticity $\langle
  e_1\rangle$ as a function of declination in the $r$ and $i$
  bands. This signal was removed from the galaxy catalogue prior to
  computing the final correlation function. The $r$ band data between
  declination $-0.8^\circ$ and $-0.4^\circ$ were rejected due to the
  known problems with camcol 2. The error bars are Poisson errors only.}
\end{figure}

The mean $e_1$ signal as a function of declination is shown in
Fig.~\ref{fig:mean_e1_plot} in bins of width 0.05 degrees.  We take
this as a template for mask-related selection biases (combined with
any systematic uncorrected PSF variation as a function of declination,
which in west-to-east drift-scan observations is a highly plausible
type of position dependence).  {\em Before computing the correlation
  function, we subtracted this mean signal from the galaxy ellipticity
  catalogue.}

\begin{figure}
\includegraphics[angle=-90,width=3.2in]{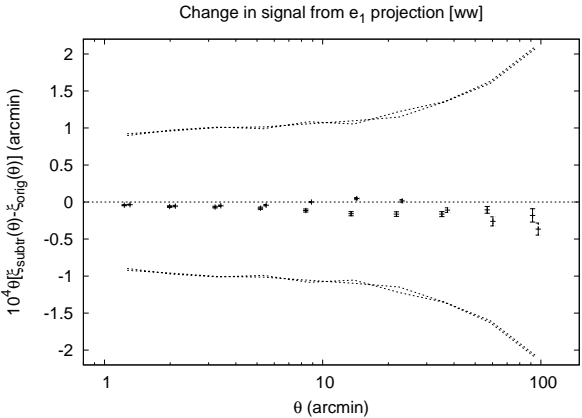}
\caption{\label{fig:e1projtest} The loss of actual power due to $e_1$ projection. Using 36 realizations from the Monte Carlo simulation, we find the difference in post-projection ellipticity correlation function $\xi(\theta)$ and original $\xi(\theta)$. These are shown as the solid points ($\xi_{++}$) and dashed points ($\xi_{\times\times}$) in the figure, re-binned to 10 bins in angular separation $\theta$. The dashed lines at top and bottom are the $\pm 1\sigma$ statistical error bars of our measurement. The reduction of actual power is detectable by combining many simulations, but is very small compared to the error bars on the measurement.}
\end{figure}

One danger in this procedure to remove spurious $\langle e_1\rangle$
is that some real power could be removed -- that is, even in the
absence of any systematic error, some of the actual galaxy shape correlation
function signal could be suppressed since the method determines the
mean $e_1$ of the real galaxies and by subtracting it introduces a
slight artificial anti-correlation. The best way to guard against this
is with simulations. Using the Monte Carlo simulation tool of
Sec.~\ref{ss:cov-mc}, we generated simulated realisations of our
ellipticity catalogue and either implemented the $\langle e_1\rangle$
projection or not. The difference in the correlation functions is a
measure of how much power was removed. The result is shown in
Fig.~\ref{fig:e1projtest}, and shows that the loss of real power is
insignificant compared to our error bars.



\subsubsection{PSF anisotropy}

Convolution with an elliptical PSF will induce a spurious ellipticity
in observed galaxy surface-brightness profiles.  While the effective
PSF for these coadds is a circular double Gaussian to quite high
precision, the tests in Paper I indicate a low level of residual
anisotropy that we must consider here.
\begin{figure*}
\includegraphics[angle=-90,width=5.5in]{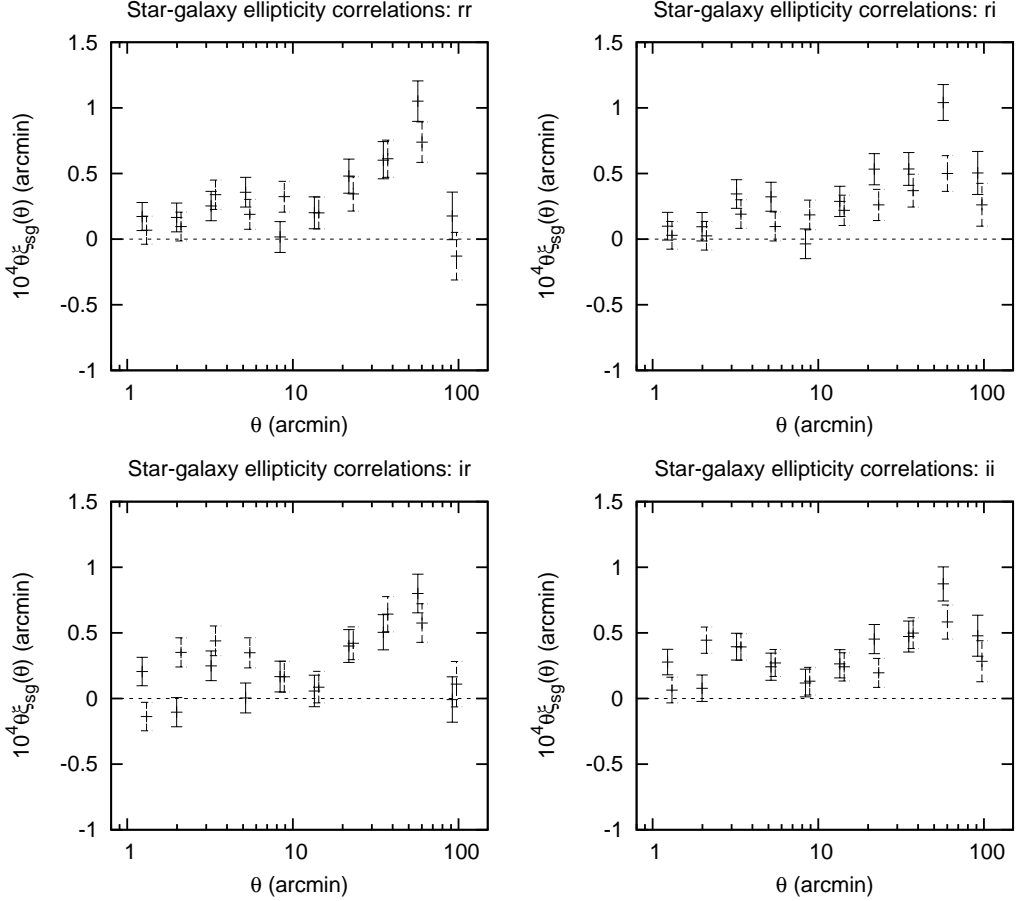}
\caption{\label{fig:sgxf}The star-galaxy ellipticity correlation
  functions. Shown are the $rr$, $ri$ (i.e. star $r$ $\times$ galaxy
  $i$), $ir$, and $ii$ correlation functions, reduced to 10 bins. The
  solid points, which are offset to slightly lower $\theta$-values for
  clarity, are the $++$ correlation functions, and the dashed points
  are the $\times\times$ functions. All error bars are Poisson only.}
\end{figure*}

Possible sources of this issue include: (i) inaccuracies in the
single-epoch PSF model used to determine the kernel to achieve the
desired PSF; (ii) colour-dependence of the PSF that means the
single-epoch PSF model from the stars is not exactly the PSF for the
galaxies; or (iii) the fact that we determine the rounding kernel on a
fixed grid, so that smaller-scale variations in PSF anisotropy might
remain uncorrected. All of these must be present at some level,
although the last two cannot be the full solution: (ii) does not
explain the residual stellar ellipticity\footnote{We have searched for
  a $g-i$ dependence in the stellar ellipticities in the coadded
  image. We only found effects at the $\sim 0.002$ level, and while
  they are statistically significant, we have not established whether
  they correspond to true colour dependence versus e.g. variation of
  stellar colour distributions along the stripe.}, and (iii) does not
explain why there is structure in the declination direction on the
scale of an entire CCD (0.23 degrees).

For a galaxy and a PSF
that are both well-approximated by a Gaussian, the PSF-correction given
above produces a measured ellipticity of:
\begin{equation}
e^\mathrm{obs} = {\cal R}_{\rm psf}e^\mathrm{PSF} = \frac{1-R_2}{R_2}e^\mathrm{PSF};
\end{equation}
see e.g. \citet{2002AJ....123..583B}.  The weighted (by the same
weights used for the correlation function; see Eq.~\ref{eqn:weight})
average of the PSF anisotropy response defined in Eq.~(\ref{eq:emodel})
over the sample of galaxies considered in this work is ${\cal R}_{\rm
  psf} = 0.86$ ($r$ band) or 0.95 ($i$ band); in what follows we take
a value of 0.9.

A nonzero star-galaxy correlation function $\xi_\mathrm{sg}$ resulting
from systematic PSF anisotropy (as estimated in Paper I) indicates the
presence of a spurious contribution to the shear-shear correlation
function with amplitude $\approx 0.9 \xi_\mathrm{sg}$. We will not
determine this response to high enough accuracy to subtract the effect
with small residual error: doing so would not require just a
simulation, but a simulation that knows the correct radial profile of
the PSF errors.\footnote{This might be an option in future space-based
  surveys if the type of error can be traced to the source of
  ellipticity (astigmatism$\times$defocus, coma, or jitter). In either
  space or ground-based data, one could imagine doing
  cross-correlations of higher-order shapelet modes
  \citep{2003MNRAS.338...35R} to extract the particular form of the
  errors. None of these options are pursued here.} In our case, the
star-galaxy correlation function is detectable but below the errors on
the galaxy-galaxy ellipticity auto-correlation (although not by very
much), so a highly accurate correction is unnecessary.

\begin{figure}
\includegraphics[angle=-90,width=3.2in]{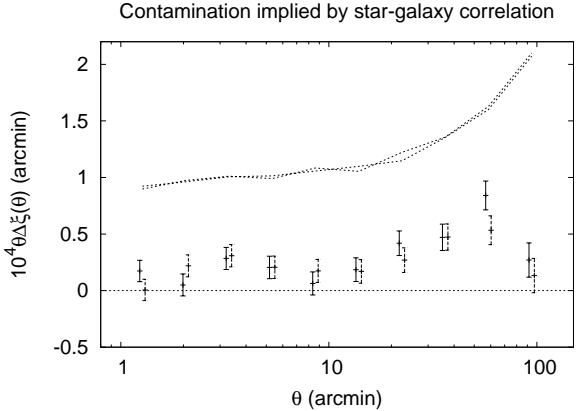}
\caption{\label{fig:sgxf_w}The implied contamination to the galaxy
  ellipticity correlation function if the star-galaxy correlation
  function is used as a measure of the additive PSF power. The solid
  points are the $++$ correlation functions, and the dashed points are
  the $\times\times$ functions. All error bars are propagated from the
  Poisson errors assuming correlation coefficient $+1$ (a better
  assumption than independent errors, but likely an overestimate). The
  dotted curves show the $1\sigma$ errors in each radial bin from the
  Monte Carlo simulations (see Sec.~\ref{ss:cov-mc}) which include
  both Poisson and cosmic variance uncertainties. Note also that the
  shapes and normalisations of the $++$ and $\times\times$ signals are
  nearly identical.}
\end{figure}

We constrain the PSF anisotropy contribution by computing the
star-galaxy correlation function. This was done in Paper I, but some
of the star-galaxy signal is due to the systematic variation of PSF
ellipticity with declination and is removed by the subtraction
procedure above. The star-galaxy ellipticity correlation function with
the corrected catalogue is shown in Fig.~\ref{fig:sgxf}. The implied
contamination to the galaxy ellipticity correlation function,
appropriately averaging the bands and applying the factor of ${\cal
  R}_{\rm psf}=0.9$, is shown in Fig.~\ref{fig:sgxf_w}.

These measured star-galaxy correlations can be used to construct a
reasonable systematics covariance matrix for this systematic. We take
the amplitude of the diagonal elements of the PSF systematic
covariance to be equal to the amplitude of the measured contamination.
We also assume that the off-diagonal terms are fully-correlated
between bins, which is equivalent to fixing the scaling of this
systematic with radius, and saying that only the overall amplitude of
the systematic is uncertain.

Since there are a number of uncertainties in this procedure, we do not
apply any correction for these additive PSF systematics as we do for
ones that are previously discussed, such as intrinsic alignments or
stellar contamination.  Instead, we simply include a term in the
systematics covariance matrix to account for it.  We also will present
a worst-case scenario for the impact of this term on cosmological
constraints; in Sec.~\ref{sec:constraints} we will show what happens
to the cosmology constraints if we assume that the systematic error is
$+2\sigma$ from its mean, i.e. $40$ per cent of the statistical
errors.  This should be taken as a worst-case scenario for this
particular systematic.

\section{Analysis tools}
\label{sec:analysis_tools}

\subsection{Ellipticity correlation function}

We compute the ellipticity correlation functions defined in
Eq.~(\ref{eq:xigamma}) on scales from 1--120 arcminutes. For the
cosmological analysis, we start by computing the correlation function
in 100 bins logarithmically spaced in separation $\theta$ to avoid bin
width artifacts. For the cosmological parameter constraints, we
project these onto the Complete Orthogonal Sets of $E$-/$B$-mode
Integrals (COSEBI) basis \citep{2010A&A...520A.116S} to avoid the
instabilities of inverting a large covariance matrix estimated via
Monte Carlo simulations (we will describe our implementation of
COSEBIs in Sec.~\ref{sec:EB-mode}). However, for display purposes, it
is more convenient to reduce the $\theta$ resolution to only 10 bins
so that the real trends are more visually apparent.

\subsubsection{Weighting}
\label{ss:weights}

The correlation functions used here are weighted by the inverse variance of the ellipticities, where the ``variance'' includes shape noise. Specifically, we define a weight for a galaxy
\begin{equation}
w_i = \frac{1}{\sigma_e^{2} + 0.37^2},
\label{eqn:weight}
\end{equation}
where $\sigma_e$ is the ellipticity uncertainty per component defined
by our shape measurement pipeline.  As demonstrated by
\cite{2011arXiv1110.4107R}, these may be significantly underestimated
in certain circumstances; however, this will only make our estimator
slightly sub-optimal, so we do not attempt to correct for it.  The
value of $0.37$ for the root-mean-square (RMS) intrinsic ellipticity
dispersion per component comes from the results of \cite{2011arXiv1110.4107R}, for
$r<22$, and therefore we are implicitly extrapolating it to fainter
magnitudes.  Given that \cite{2007ApJS..172..219L} found a constant
RMS ellipticity to far fainter magnitudes in the COSMOS data, we
consider this extrapolation justified\footnote{Note that we do not use
  the actual value of RMS ellipticity from \cite{2007ApJS..172..219L}
  -- only the trend with magnitude -- because, as demonstrated by
  \cite{2011arXiv1107.4629M}, the RMS ellipticity value in
  \cite{2007ApJS..172..219L} is not valid for our adaptively-defined
  moments, which use an elliptical weight function matched to the
  galaxy light profile.}.

\begin{figure*}
\includegraphics[angle=-90,width=6.5in]{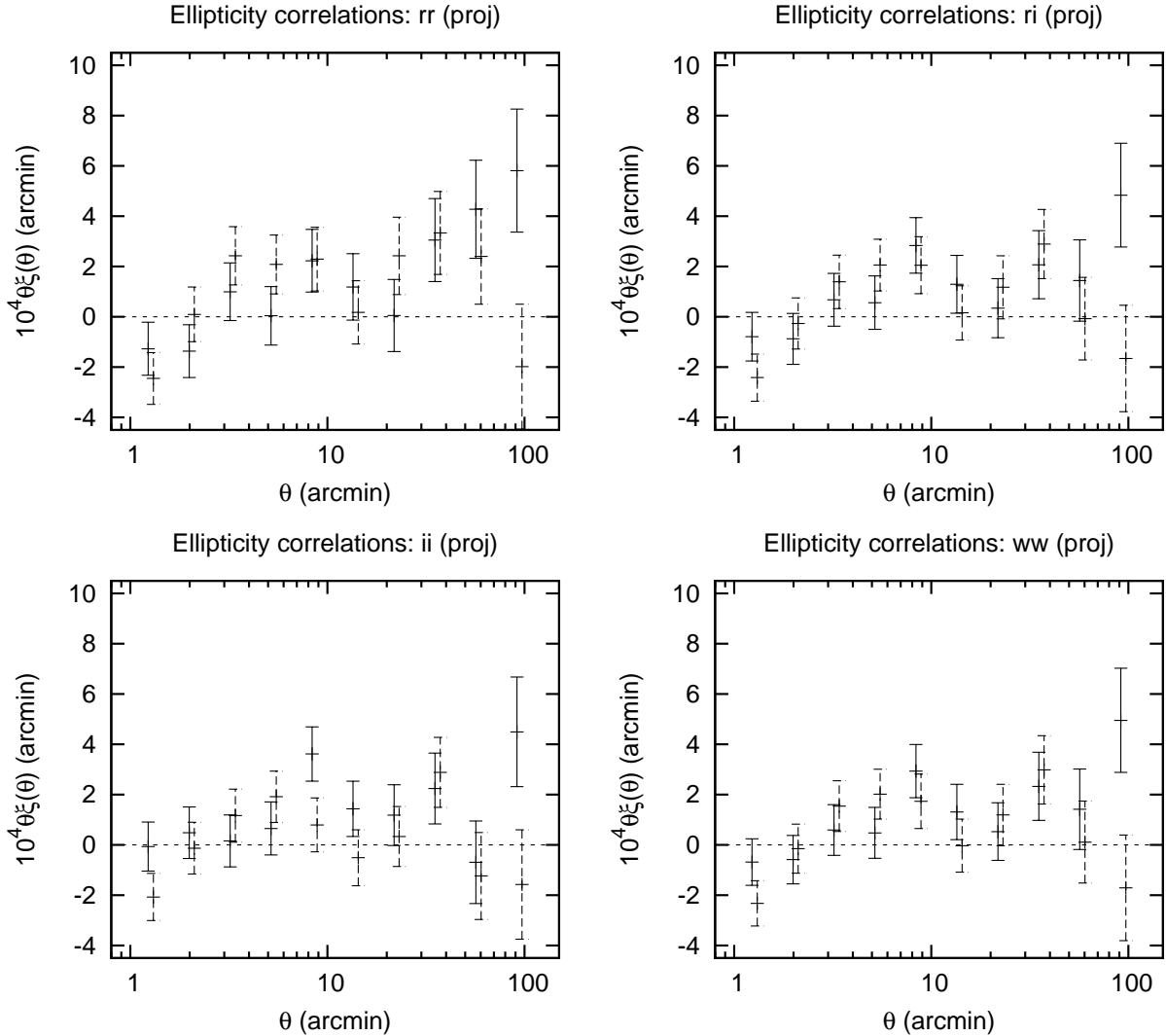}
  \caption{\label{fig:cfplotproj}The ellipticity correlation functions in
    the $rr$, $ri$, $ii$ and $ww$ (combined) band combinations. The
    solid points denote the $++$ and the dashed points denote the
    $\times\times$ components of the correlation function. The points
    have been slightly displaced horizontally for clarity. The Monte
    Carlo errors are shown.}
\end{figure*}

\subsubsection{Direct pair-count code}
\label{ss:cf-pcc}

A direct pair-count correlation function code was used for the
cosmological analysis. It is slow ($\sim 3$ hours for $2\times 10^6$
galaxies on a modern laptop) but robust and well-adapted to the Stripe
82 survey geometry. The code sorts the galaxies in order of increasing
right ascension $\alpha$; the galaxies are assigned to the range
$-60^\circ<\alpha<+60^\circ$ to avoid unphysical edge effects near
$\alpha=0$. It then loops over all pairs with
$|\alpha_1-\alpha_2|<\theta_{\rm max}$. The usual ellipticity
correlation functions can be computed, e.g.
\begin{equation}
\xi_{++}(\theta) = \frac{\sum_{ij} w_i w_j e_{i+} e_{j+}}{\sum_{ij} w_i w_j},
\end{equation}
where the sum is over pairs with separations in the relevant $\theta$
bin, and the ellipticity components are rotated to the line connecting
the galaxies. The direct pair-count code works on a flat sky,
i.e. equatorial coordinates $(\alpha,\delta)$ are approximated as
Cartesian coordinates. This is appropriate in the range considered,
$|\delta|<1.274^\circ$, where the maximum distance distortions are
$\frac12\delta^2_{\rm max} = 2.5\times 10^{-4}$.  The direct
pair-count code is applicable to either auto-correlations of galaxy
shapes measured in a single filter ($rr$, $ii$) or cross-correlations
between filters or between distinct populations of objects ($ri$ and
all of the star-galaxy correlations).

Simple post-processing allows one to compute the $\xi_+$ and $\xi_-$
correlation functions, defined by
\begin{equation}
\xi_+(\theta) \equiv \xi_{++}(\theta)+\xi_{\times\times}(\theta)
\end{equation}
and
\begin{equation}
\xi_-(\theta) \equiv \xi_{++}(\theta)-\xi_{\times\times}(\theta).
\end{equation}

\subsubsection{Combining bands}
\label{ss:combo}

Finally, the different band correlation functions $rr$, $ri$, and $ii$
must be combined according to some weighting scheme:
\begin{equation}\label{eq:avgellipcorr}
\xi^{ww}_{++}(\theta) = w_{rr}\xi^{rr}_{++}(\theta) + w_{ri}\xi^{ri}_{++}(\theta) + w_{ii}\xi^{ii}_{++}(\theta),
\end{equation}
where the label ``$ww$'' indicates that the bands were combined. The
relative weights were chosen according to the fraction of measured
shapes in $r$- and $i$-bands, i.e. $w_{rr}=f_r^2$, $w_{ri}=2f_rf_i$,
and $w_{ii}=f_i^2$ where the weights are $f_r = 0.4603$ and $f_i =
0.5397$.

The final ellipticity correlation functions (with the $\theta$
resolution reduced to 10 bins) are shown in Fig.~\ref{fig:cfplotproj}.

\subsection{Tests of the correlation function}
\label{ss:tests}

We implement several null tests on the correlation function to search
for remaining systematic errors.

\begin{figure}
\includegraphics[angle=-90,width=3.2in]{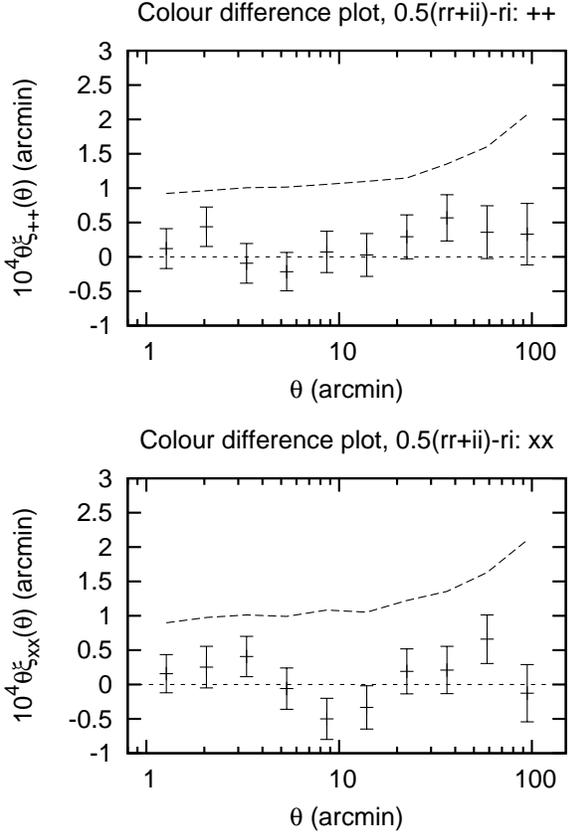}
\caption{\label{fig:nulplot}The difference between the galaxy
  ellipticity cross-correlations ($ri$) and the auto-correlations
  $(rr+ii)/2$, with error bars determined from the Monte Carlo
  simulations. The upper panel shows the $++$ correlations and the
  lower panel shows the $\times\times$ correlations. The dashed line
  is the $1\sigma$ statistical error bar on the actual
  signal.}
\end{figure}

The first test, shown in Fig.~\ref{fig:nulplot}, constructs the
difference between the cross-correlation function of $r$ and $i$ band
galaxy ellipticities versus the $rr$ and $ii$ auto-correlations. The
differences in the two types of correlation functions are small
compared to the statistical uncertainty in the signal. This is
consistent with our expectations, as the true cosmic shear signal
should be independent of the filters in which galaxy shapes are
measured.

\begin{figure*}
\begin{center}
\includegraphics[angle=-90,width=6.5in]{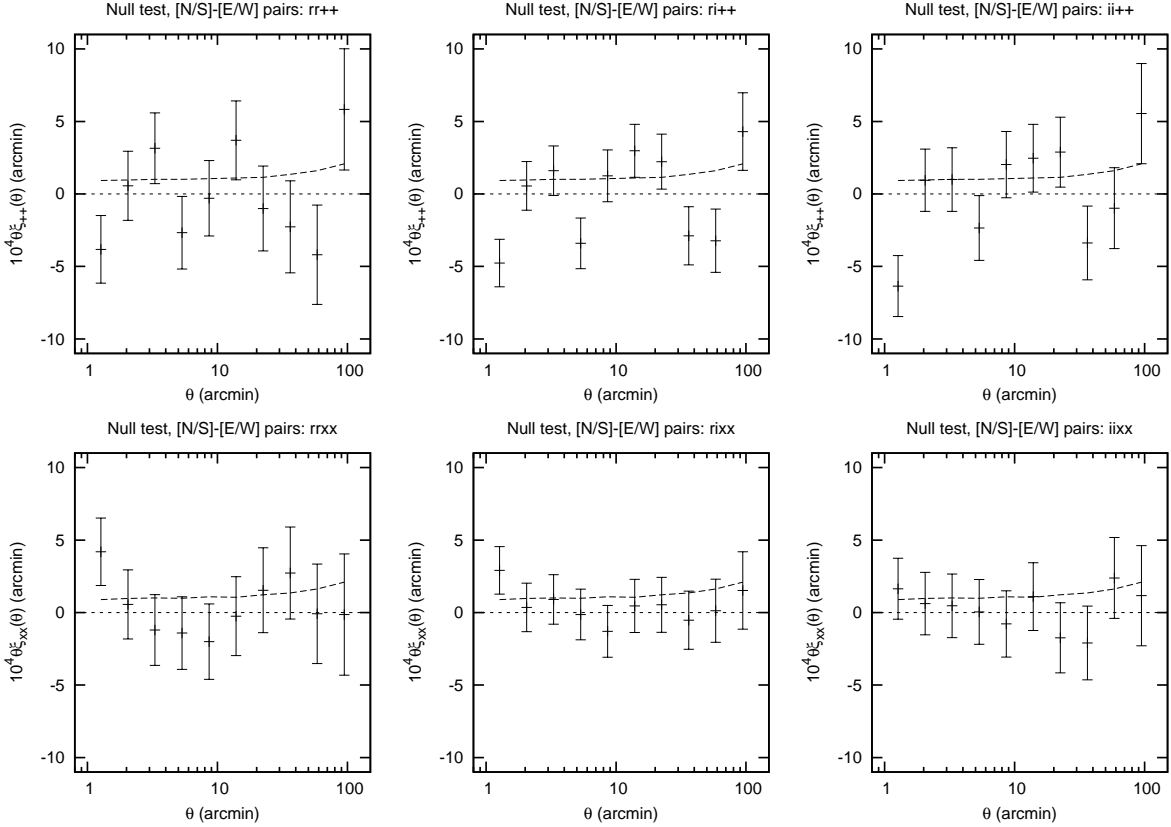}
\caption{\label{fig:null_ns_ew}The null test of the correlation functions measured using galaxy pairs whose separation vector is within 45$^\circ$ of the north-south direction, minus that measured using galaxy pairs whose separation vector is within 45$^\circ$ of the east-west direction. The error bars shown are the Poisson errors only. The dashed curve shows the $1\sigma$ error bars of the actual signal (all colour combinations and separation vectors averaged). The 6 panels show the three colour combinations ($rr$, $ri$, and $ii$) and the 2 components ($++$ or $\times\times$).}
\end{center}
\end{figure*}

The second test, shown in Fig.~\ref{fig:null_ns_ew}, compares the
(band averaged or $ww$) correlation function computed using galaxy
pairs separated in the cross-scan (north-south) direction versus pairs
separated in the along-scan (east-west) direction. This difference
should be zero if the signal we measure is due to lensing in a
statistically isotropic universe. The error bars shown are Poisson
errors, so they may be slight underestimates at the larger scales,
where cosmic variance becomes important. Visual inspection shows no
obvious offset from zero, but the error bars are larger for this test
than in Fig.~\ref{fig:nulplot} because the null test includes no
cancellation of galaxy shape noise.

\subsection{E/B-mode decomposition}
\label{sec:EB-mode}

As a final check for systematics, we decompose the 2-point correlation
function (2PCF) into $E$- and $B$-modes, where, to leading order,
gravitational lensing only creates $E$-modes. The $B$-modes can arise
from the limited validity of the Born approximation
\citep{jsw00,hhw09}, redshift source clustering
\citep{2002A&A...389..729S}, and lensing (magnification) bias
\citep{srd09,krh10}, however the amplitude of $B$-modes from these
sources should be undetectable with our data. At our level of
significance, a $B$-mode detection would indicate remaining
systematics, e.g. due to spurious power from an incomplete PSF
correction.

Formerly used methods to decompose $E$-and $B$-modes, such as the aperture
mass dispersion 
\begin{equation}
\label{eq:map}
 \langle M_\mr{ap}^2\rangle (\theta)  =  \int_0^{2\theta} \frac{\mr d \vartheta \, \vartheta}{2 \, \theta^2} \left[ \xi_+ (\vartheta) T_+ \left( \frac{\vartheta}{\theta} \right)\; + \; \xi_- (\vartheta) T_- \left( \frac{\vartheta}{\theta} \right)\right], 
\end{equation} 
with the filter functions $T_\pm$ as derived in
\cite{2002A&A...389..729S}, or the shear $E$-mode correlation
function, suffer from $E/B$-mode mixing \citep{2006A&A...457...15K},
i.e. $B$-modes affect the $E$-mode signal and vice versa. These
statistics can be obtained from the measured 2PCF, for an exact
$E/B$-mode decomposition, however they require information on scales
outside the interval $[\tmin;\tmax]$ for which the 2PCF has been
measured.

\begin{figure}
\includegraphics[width=7.0cm]{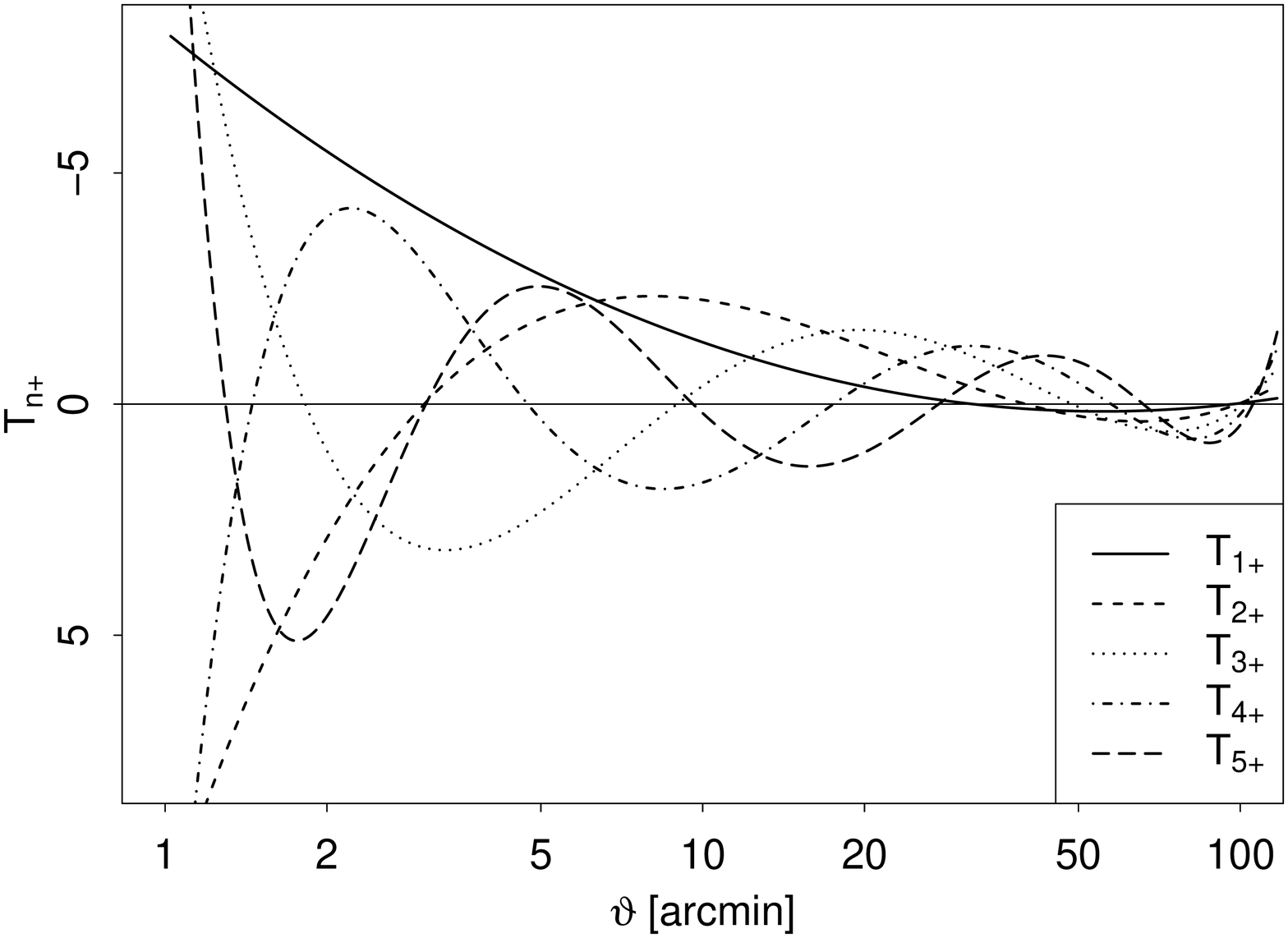}
\includegraphics[width=7.0cm]{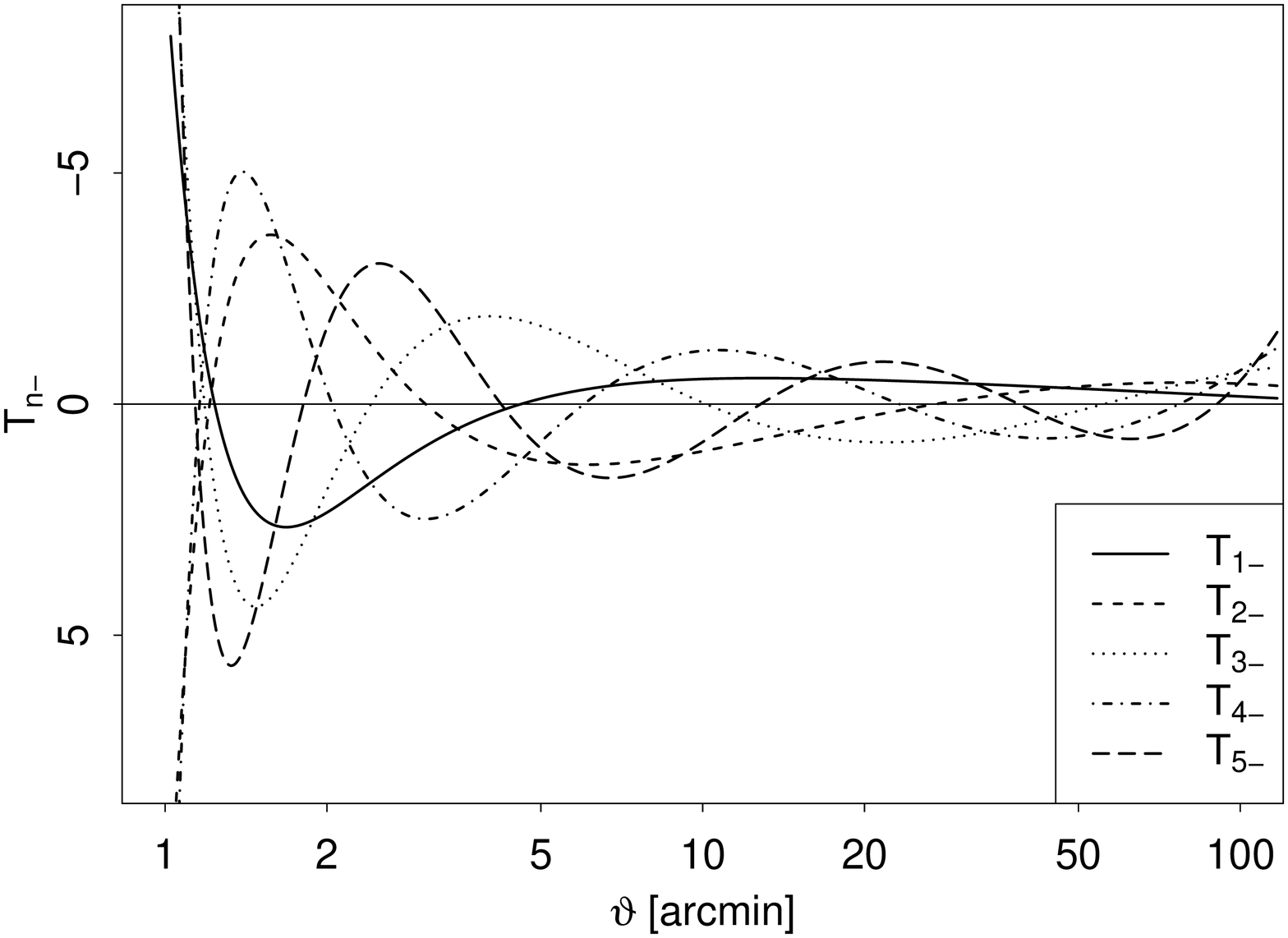}
\caption{\label{fig:cosebi-t} The COSEBI filter functions $T_{n+}$ (upper panel) and $T_{n-}$ (lower panel) for the first 5 modes.}
\end{figure}

The ring statistics
\citep{sck07,2010A&A...510A...7E,2010MNRAS.401.1264F} and more
recently the COSEBIs \citep{2010A&A...520A.116S} perform an EB-mode
decomposition using a 2PCF measured over a finite angular
range. COSEBIs and ring statistics can be expressed as integrals over
the 2PCF as
\begin{equation}
EB = \int_{\tmin}^{\tmax} \frac{{\rm d}\vt}{2}\;\vt[\tplog(\vt) \xi_+(\vt) \pm \tmlog(\vt) \xi_-(\vt)]
\label{eq:cosebis}
\end{equation}
and
\begin{equation}
R_{\rm{EB}} (\vt) = \int_{\vt_{\rm min}}^\vt  \frac{\rm d \vt'}{2\vt'}[ \xip(\vt') Z_+(\vt',\vt) \pm \xim(\vt') Z_-(\vt',\vt)].
\label{eq:ring}
\end{equation}
For the ring statistics, we use the filter functions $Z_\pm$ specified
in \cite{2010A&A...510A...7E}. The derivation of the COSEBI filter
functions $T_{\pm n}$ is outlined in \cite{2010A&A...520A.116S}, where
the authors provide linear and logarithmic filter functions indicating
whether the separation of the roots of the filter function is
distributed linearly or logarithmically in $\vt$. Note that whereas
the ring statistics are a function of angular scale, the COSEBIs are
calculated over the total angular range of the 2PCF, condensing the
information from the 2PCF naturally into a set of discrete modes.  The
linear $T$-functions can be expressed conveniently as Legendre
polynomials, however $\tpmlog$ compresses the cosmological information
into significantly fewer modes; we therefore choose the logarithmic
COSEBIs as our second-order shear statistic in the likelihood analysis
in Sec.~\ref{sec:constraints}. The COSEBI filter functions are
displayed graphically in Fig.~\ref{fig:cosebi-t}.

\begin{figure*}
\includegraphics[width=5.5cm]{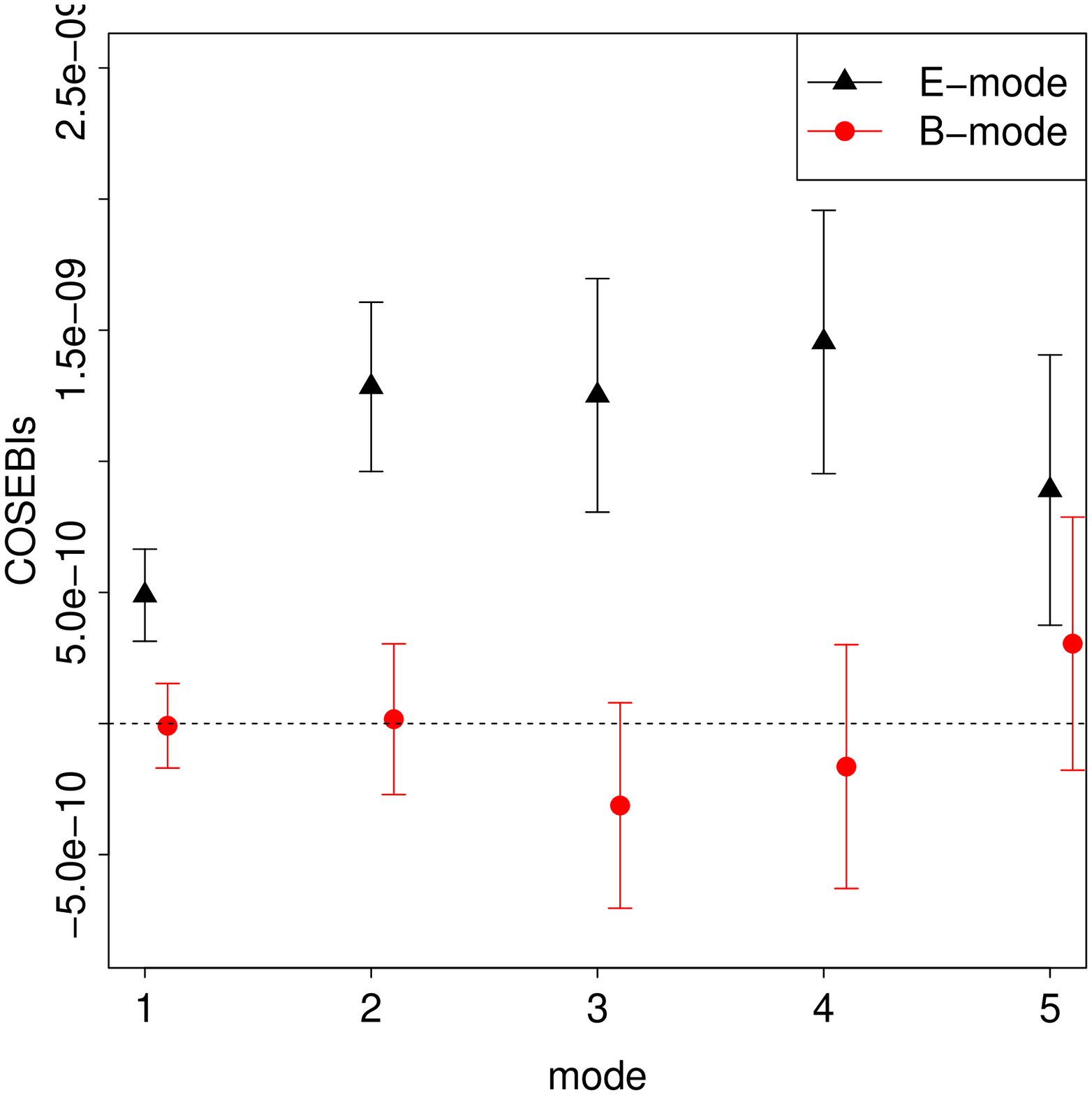}
\includegraphics[width=5.5cm]{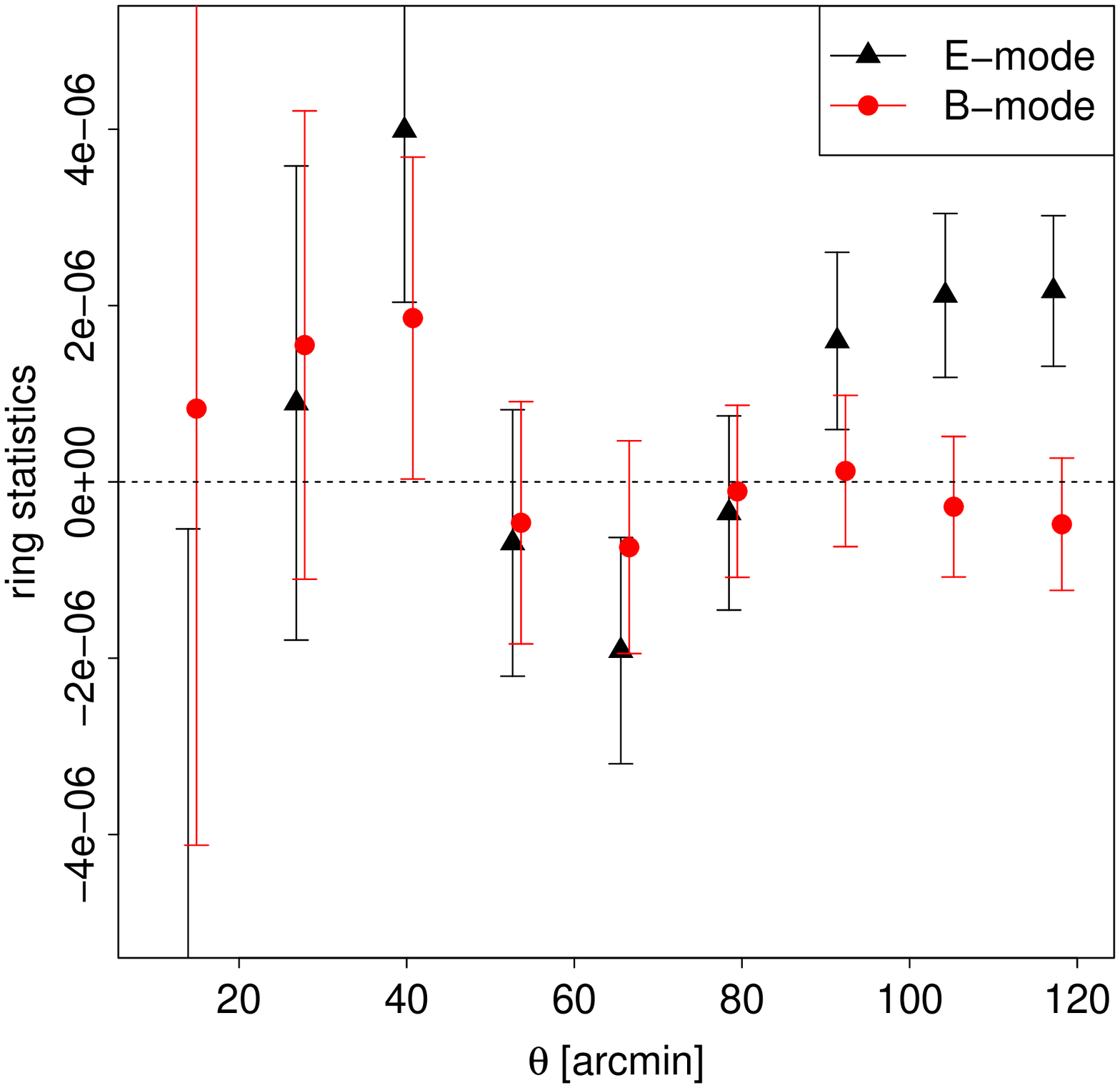}
\includegraphics[width=5.5cm]{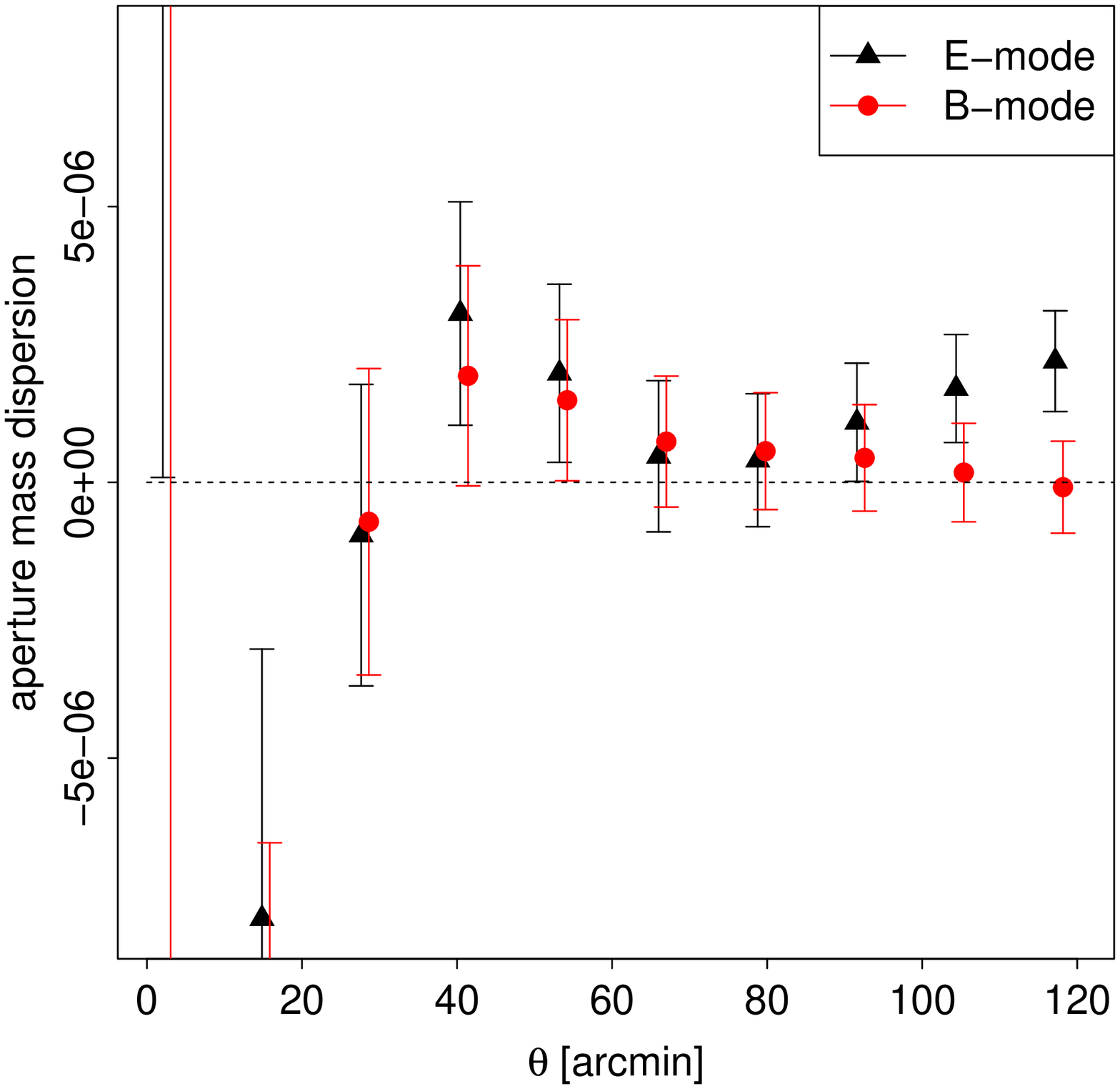}
\caption{\label{fig:eb-mode} The measured COSEBIs, ring statistics, and  aperture mass dispersion from the combined cosmic shear  signal. The error bars equal the square root of the corresponding covariances' diagonal elements (statistics only). Note that the COSEBIs data points are significantly correlated. Slightly smaller is the correlation for the aperture mass dispersion, and the ring statistics' data points have the smallest correlation.}
\end{figure*}

Figure \ref{fig:eb-mode} shows three different $E/B$-mode statistics derived
from our measured shear-shear correlation function, i.e. the COSEBIs,
the ring statistics, and the aperture mass dispersion. The error bars are  obtained from the square root of the corresponding covariances' diagonal elements (statistics only). Note that the COSEBIs data points are significantly correlated. Slightly smaller is the correlation for the aperture mass dispersion, and the ring statistics' data points have the smallest correlation. 

From the COSEBIs, we find a reduced $\chi^2$ for the $E$-modes to be consistent with zero of 6.395, versus 1.096 for the $B$-modes (5 degrees of freedom each). The latter is consistent with purely statistical fluctuations.

\section{Covariance estimation}\label{sec:allcov}

\subsection{Ellipticity correlation function covariance matrix}
\label{sec:cov}

The covariance matrix of the ellipticity correlation function
estimated via Eq.~(\ref{eq:avgellipcorr}) was computed in
several ways. The preferred method for our analysis is a Monte Carlo
method (Sec.~\ref{ss:cov-mc}) but we compare that covariance matrix 
with an estimate of the Poisson errors (Sec.~\ref{ss:poisson}) as a consistency check. 

\subsubsection{Poisson method}
\label{ss:poisson}

The direct pair-count correlation function code can compute
the Poisson error bars, i.e. the error bars neglecting the
correlations in $e_{i+}e_{j+}$ between different pairs. This estimate
of the error bar is
\begin{equation}
\sigma^2[\xi_{++}(\theta)] = \frac{\sum_{ij} w_i^2 w_j^2 |{\bmath e}_i|^2 |{\bmath e}_j|^2}{2\left[ \sum_{ij} w_i w_j \right]^2}.
\end{equation}
Equivalently, this is the variance in the correlation function that one
would estimate if one randomly re-oriented all of the galaxies. The
Poisson method is simple, however it is not fully appropriate for $ri$
cross-correlations (since the same intrinsic shape noise is recovered
twice for pairs that appear in both $ri$ and $ir$
cross-correlations). Moreover, at scales of tens of arcminutes and
greater there is an additional contribution because the cosmic shear
itself is correlated between pairs. Therefore the Poisson error bars
should be used only as a visual guide: they would underestimate the
true uncertainties if used in a cosmological parameter analysis.

\subsubsection{Monte Carlo method}
\label{ss:cov-mc}

We used a Monte Carlo method to compute the covariance matrix of
$\xi_{++}(\theta)$ and $\xi_{\times\times}(\theta)$. The method is
part theoretical and part empirical: it is based on a theoretical
shear power spectrum, but randomizes the real galaxies to correctly
treat the noise properties of the survey. The advantages of the Monte
Carlo method -- as implemented here -- are that spatially variable
noise, intrinsic shape noise including correlations between the $r$
and $i$ band, and the survey window function are correctly
represented. The principal disadvantages are that the cosmic shear
field is treated as Gaussian and a particular cosmology must be
assumed \citep[see][for alternative approaches]{2009A&A...502..721E}. However, so long as this cosmology is not too far from the
correct one (an assumption that can itself be tested!), the Monte Carlo
approach is likely to yield the best covariance matrix.

The Monte Carlo approach begins with the generation of a suite of $\NMC$
realizations of a cosmic shear field in harmonic space according to a
theoretical spectrum. For our analysis, the theoretical spectrum was
that from the {\slshape WMAP} 7-year \citep{2011ApJS..192...16L} cosmological parameter set
(flat $\Lambda$CDM; $\Omega_{\rm b}h^2
= 0.02258$; $\Omega_{\rm m}h^2 = 0.1334$; $n_{\rm s}=0.963$;
$H_0=71.0\,$km$\,$s$^{-1}\,$Mpc$^{-1}$; and $\sigma_8=0.801$), and the
shear power spectrum code used in \citet{2009arXiv0901.0721A}, itself
based on the \citet{1998ApJ...496..605E} transfer function and the
\citet{2003MNRAS.341.1311S} nonlinear mapping. The redshift
distribution discussed in section \ref{sss:zdist}, based on a
calibration sample from DEEP2, VVDS, and PRIMUS, was used as the input
to the shear power spectrum calculation.

From this power spectrum we generate a sample set of Gaussian $E$-mode
shear harmonic space coefficients $a^{\rm E}_{lm}$. The full power
spectrum is used at $l\le1500$; a smooth cutoff is applied from
$1500<l<2000$ and no power at $l\ge 2000$ is included. This is
appropriate for a covariance matrix since the power at smaller scales
is shot noise dominated and cannot be recovered. (The $E$-mode power
spectrum is $C^{EE}_{1500} = 3.6\times 10^{-11}$, as compared to a
shot noise of $\gamma_{\rm int}^2/\bar n \sim 1.8\times 10^{-9}$.)  No
$B$-mode shear is included. The particle-mesh spherical harmonic
transform code of \citet{2004PhRvD..70j3501H} with a $6144\times 3072$
grid ($L'=6144$) and a 400-node interpolation kernel ($K=10$) was used
to transform these coefficients into shear components
$(\gamma_1,\gamma_2)$ at the position $\hat{\bmath n}_j$ of each
galaxy $j$.\footnote{The use of a full-sky approach for the Monte
  Carlo realisations was not necessary for the SDSS Stripe 82 project,
  but was the simplest choice given legacy codes available to us.}

A synthetic ellipticity catalogue was then generated as follows. For
each galaxy, we generated a random position angle offset
$\psi_j\in[0,\pi)$ and rotated the ellipticity in both $r$ and $i$
bands by $\psi_j$.\footnote{To simplify bookkeeping, the actual
  implementation was that a sequence of $10^7$ random numbers was
  generated, and a galaxy was assigned one of these numbers based on
  its coordinates in a fine grid with 0.36 arcsec cells in
  $(\alpha,\delta)$.} We then added the synthetic shear weighted by
the shear responsivity to the randomised ellipticity to generate a
synthetic ellipticity:
\begin{equation} 
{\bmath e}^{\rm syn}_j = \rme^{2\rmi \psi_j} {\bmath
    e}^{\rm true}_j + 1.73 \bgamma(\hat{\bmath n}_j).
\end{equation}
The 1.73 prefactor was estimated from Eq.~(\ref{eq:perfectcal}), which
we expected to be good enough for use in the Monte Carlo analysis, so
that the Monte Carlos could be run in parallel with the shear
calibration simulations. The latter gave a final result of
$1.78\pm0.04$, which is not significantly different.

The direct pair-count correlation function code, in all versions
($rr$, $ri$, and $ii$) was run on each of the $\NMC$ Monte Carlo
realisations, before combining the different correlations to get the
weighted value via Eq.~(\ref{eq:avgellipcorr}).

\begin{figure}
\includegraphics[angle=-90,width=3.2in]{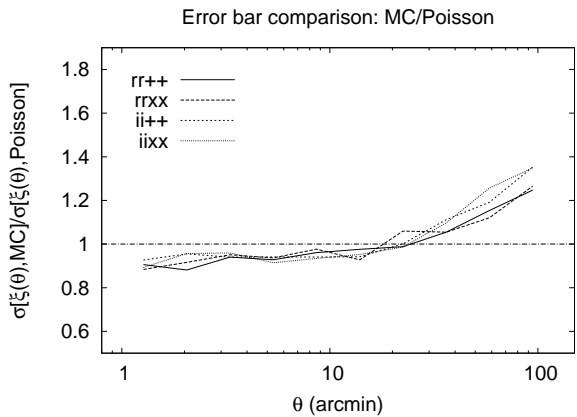}
  \caption{\label{fig:er}The ratio of error bars obtained by the Monte
    Carlo method to those obtained by the Poisson method, for 10
    angular bins. The four curves show either $rr$ or $ii$ band
    correlation functions, and either the $++$ or $\times\times$
    component. Note the rise in the error bars at large values of
    the angular separation, due to mode sampling variance.}
\end{figure}

\begin{figure}
\includegraphics[angle=-90,width=3.2in]{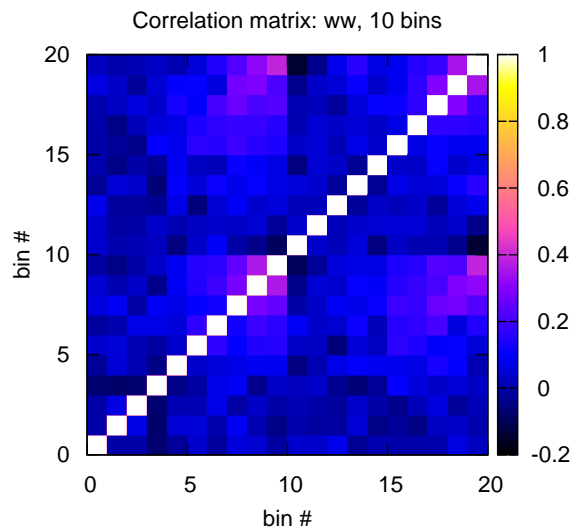}
\caption{\label{fig:corrplot.ww.R}The matrix of correlation
  coefficients for the combined ($ww$) correlation functions in the 10
  angular bins for which the correlation function is plotted in the
  companion figures. The bin number ranges from 0--9 for
  $\xi_{++}(\theta)$ and from 10--19 for $\xi_{\times\times}(\theta)$;
  all diagonal components are by definition equal to unity. Based on
  \NMC\ Monte Carlo realisations.}
\end{figure}

The Monte Carlo and Poisson error bars are compared in
Fig.~\ref{fig:er}. The correlation coefficients of the correlation
functions in different bins are plotted graphically in
Fig.~\ref{fig:corrplot.ww.R}.

\begin{figure}
\includegraphics[width=8cm]{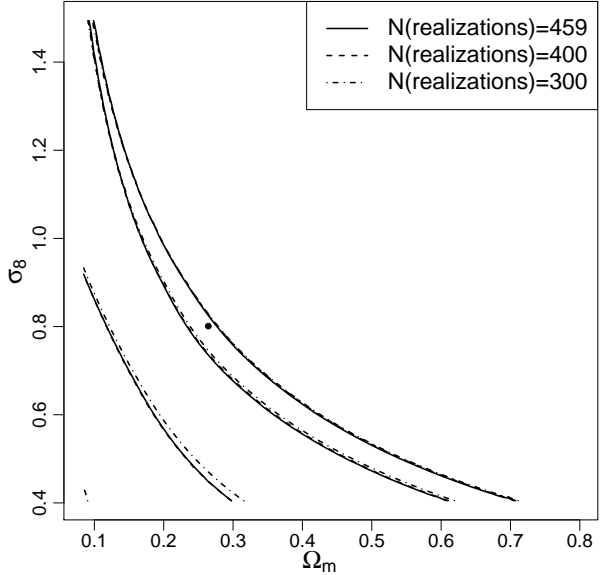}
\caption{\label{fig:mc_converge}Convergence test of the $\sigma_8$
  vs. $\Omega_m$ parameter constraints as a function of the number of
  Monte Carlo realizations used to compute the covariance. The plot
  shows the $68$ and $95$ per cent likelihood contours (however, the
  lower 95 per cent contours are not visible). The covariance includes
  statistical errors only.}
\end{figure}

From each Monte Carlo correlation function we compute the COSEBIs via
Eq.~(\ref{eq:cosebis}) and use their covariance matrix in our
subsequent likelihood analysis. In order to test whether our
covariance has converged, meaning that the number of realisations is
sufficient to not alter cosmological constraints, we perform 3
likelihood analyses in $\sigma_8$ vs. $\Omega_m$ space varying the
numbers of realisations from which we compute the covariance matrix
(see Sec.~\ref{sec:constraints} for detailed methodology; for now we
are just establishing convergence of the covariance matrix). In
Fig.~\ref{fig:mc_converge} we show the $68$ and $95$ per cent
likelihood contours, i.e. the contours enclose the corresponding
fraction of the posterior probability (within the ranges of the
parameters shown). We see that the contours hardly change when going
from 300 to 400 realizations and show no change at all when going from
400 to 459 realisations, hence the 459 Monte Carlo realizations are
sufficient for our likelihood analysis.


\subsection{Systematic contributions to the covariance matrix}\label{subsec:syscontrib}

The following additional contributions are added to the Monte Carlo
covariance matrix (and if appropriate the theory result) described in
Sec.~\ref{ss:cov-mc}.
\begin{enumerate}
\item The intrinsic alignment error was included following
  Sec.~\ref{subsec:ia}: the theory shear correlation function was
  reduced by a factor of 0.92, and an uncertainty of 4 per cent of the
  theory was added to the covariance matrix, i.e. we add an intrinsic
  alignment contribution
\begin{equation}
{\rm Cov}[\xi_i,\xi_j]({\rm intrinsic~alignment)} = 0.04^2 \xi_i^{\rm(th)}\xi_j^{\rm(th)},
\end{equation}
where the theory curve (th) is obtained at the fiducial WMAP7 point.
This covariance matrix includes perfect correlation between radial
bins, implying that we treat this systematic as being an effect with a
fixed scaling with separation, so the only degree of freedom is its
amplitude.
\item The stellar contamination was included following
  Sec.~\ref{ss:stellar}: the theory shear correlation function was
  reduced by a factor of 0.936, and an uncertainty of 3 per cent of
  the theory was added to the covariance matrix, i.e. we add a stellar
  contamination contribution
\begin{equation}
{\rm Cov}[\xi_i,\xi_j]({\rm stellar~contamination)} = 0.03^2 \xi_i^{\rm(th)}\xi_j^{\rm(th)},
\end{equation}
where the theory curve (th) is obtained at the fiducial WMAP7 point.
\item
The implied error from the redshift distribution uncertainty is derived from 402 realisations
of the sampling variance simulations as described in Sec.~\ref{sss:z-uncert}. We construct the
covariance matrix of the predicted $E$-mode COSEBIs.
\item
The shear calibration uncertainty was conservatively estimated in Sec.~\ref{subsec:shearcalib} to be $\pm 2.4$ per cent, or equivalently 4.8 per cent in second-order statistics. We thus add another term to the covariance matrix,
\begin{equation}
{\rm Cov}[\xi_i,\xi_j]({\rm shear~calibration)} = 0.048^2 \xi_i^{\rm(th)}\xi_j^{\rm(th)}.
\end{equation}
\item In Sec.~\ref{subsec:additive}, we described a procedure for
  including uncertainty due to additive PSF contamination.  According
  to this procedure, the relevant systematics covariance matrix is
  related to the amplitude of the measured contamination signal:
\begin{equation} 
  {\rm Cov}[\xi_i, \xi_j]({\rm PSF~contamination)} =  0.9^2 \xi_{{\rm sg},i} \xi_{{\rm sg},j},
\end{equation}
again assuming a fixed scaling with radius for this systematic
uncertainty. Since all entries scale together, we do not spuriously
``average down'' our estimate of the systematic error by combining
many bins.
\end{enumerate}

The final data vector and its covariance matrix (including all the
statistical and systematic components) are given in
Tables~\ref{table:datavector} and~\ref{table:covar}.  Note that given our
procedure of applying the systematic corrections to the theory, the
data vector is the observed one without any such corrections for the
stellar contamination and intrinsic alignments contamination.
With this
in hand, we can estimate the significance of the $E$- and $B$-mode
signals 
described in section \ref{sec:EB-mode}. The probability that the
COSEBI $E$-mode signal that we observe is due to random chance given the null
hypothesis (no cosmic shear) is
$6.0\times 10^{-6}$. The
probability of measuring our $B$-mode signal due to random chance
given the null hypothesis of zero $B$ modes is $.36$, evidence that there is no
significant B-mode power.

\section{Cosmological Constraints}
\label{sec:constraints}

Having described the measured cosmic shear two-point statistics, and
shown that the systematic bias in this measurement is small compared
with the statistical constraints, we now turn to the cosmological
interpretation. We work in the context of the flat $\Lambda$CDM
parametrisation, taking where necessary the WMAP7
\citep{2011ApJS..192...18K} constraints for our fiducial parameter
values.

\subsection{The prediction code: modeling second-order shear statistics}
\label{sec:predictions}

To produce a cosmological interpretation of our measured cosmic shear
signal from our model framework, we require a method to convert a
vector of cosmological parameters into a prediction of the observed
cosmic shear signal. Due to projection effects, we expect that a
significant fraction of the observed cosmic shear signal is produced
by the clustering of matter on nonlinear scales, so a suitably
accurate prediction algorithm must ultimately rely on numerical
simulations of structure formation.

The prediction code used in our likelihood analysis is a modified
version of the code described in \cite{eif11}. We combine Halofit
\citep{2003MNRAS.341.1311S}, an analytic approach to modeling
nonlinear structure, with the Coyote Universe Emulator \citep{lhw10},
which interpolates the results of a large suite of high-resolution
cosmological simulations over a limited parameter space, to obtain the
density power spectrum. The derivation is a two-step process: First,
we calculate the linear power spectrum from an initial power law
spectrum $P_{\delta}(k) \propto k^{n_{\rm s}}$ employing the dewiggled
transfer function of \cite{1998ApJ...496..605E}.  The non-linear
evolution of the density field is incorporated using Halofit. In order
to simulate $w$CDM models we follow the scheme implemented in {\sc
  icosmo} \citep{rak11}, interpolating between flat and open
cosmological models to mimic Quintessence cosmologies (see
\citealt{2010A&A...516A..63S} for more details).  In a second step, we
match the Halofit power spectrum to the Coyote Universe emulator
(version 1.1) power spectrum, which emulates $\pd$ over the range
$0.002 \le k \le 3.4h/$Mpc within $0\le z\le 1$ to an accuracy of 1
per cent. Wherever possible, the matched power spectrum exactly
corresponds to the Coyote Universe Emulator; of course this is limited
by the cosmological parameter space of the Emulator and its limited
range in $k$ and $z$.  However, even outside the range of the
Emulator, we rescale the Halofit power spectrum with a scale factor
$\pd$(Coyote)/$\pd$(Halofit) calculated at the closest point in
parameter space (cosmological parameters, $k$, and $z$) where the
Emulator gives results. Outside the range of the Emulator, the
accuracy of this ``Hybrid'' density power spectrum is of course worse
than 1 per cent, however it should be a significant improvement over a
density power spectrum from Halofit only. From the so-derived density
power spectrum we calculate the shear power spectrum via
Eq.~(\ref{eq:pkappa}) and the shear-shear correlation function via
Eq.~(\ref{eq:xigamma}). As a final step, we transform these predicted
correlation functions to the COSEBI basis as described above in
Sec.~\ref{sec:EB-mode}.

\begin{figure}
\includegraphics[width=7cm]{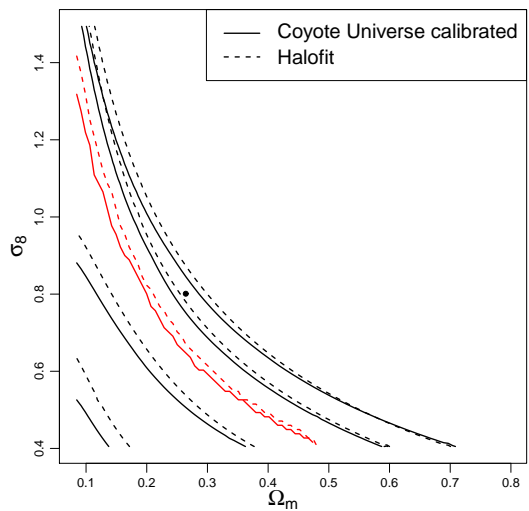}
\caption{\label{fig:coyote}The 68 and 95 per cent likelihood contours of the
combined data vector including a full treatment of systematics
when using the Halofit prediction code (dashed) and when using
the Coyote Universe-calibrated prediction code (solid). The red
lines correspond to the best-fitting value of $\sigma_8$ for a given $\Omega_m$. The dot indicates
  the WMAP7 best-fitting values.}
\end{figure}

For our final results in the $(\Omega_m,\sigma_8)$ likelihood analysis,
we used both prediction codes; the results are
compared in Fig.~\ref{fig:coyote}, where they are seen to agree to
much better than $1\sigma$.  We therefore conclude that uncertainty in
the theory predictions is sub-dominant to the other sources of
systematic error, and to the statistical error.


\subsection{Constructing the input data vector}
\label{sec:input data vector}

\begin{figure}
\includegraphics[width=7cm]{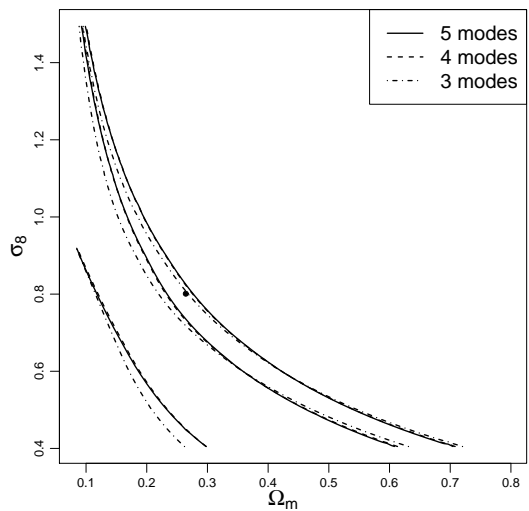}
\caption{Convergence test of the $\sigma_8$ vs. $\Omega_m$ parameter
  constraints as a function of number of COSEBI modes in the data
  vector. The plot shows the likelihood contours enclosing $68$ and
  $95$ per cent of the posterior distribution.  (The lower bounding
  curve for the 95 per cent contours is not visible on the plot.)  The
  covariance contains statistical errors only. The dot indicates the
  WMAP7 best-fitting values.\label{fig:mode_convergence} }
\end{figure}

For our primary science results, we use the measured 5 COSEBI modes
(see Fig. \ref{fig:eb-mode}, left panel). As a first step we want to
determine the number of COSEBI modes that need to be included in our
likelihood analysis. In Fig. \ref{fig:mode_convergence} we show a
likelihood analysis in the $\sigma_8$-$\Omega_{\rm m}$ parameter space
varying the number of modes in the data vector. We find that there is
hardly a change in the likelihood contours when going from 4 to 5
modes; we therefore conclude that 5 modes is a sufficient number to
capture the cosmological information encoded in our data set.

As shown in \cite{eks08}, the information content of the aperture mass
dispersion can be greatly improved when including 1 data point of the
shear-shear correlation function $\xi_+$ into the data vector; here we
adopt this concept for the COSEBIs. The basic idea is that the data
point of the correlation function is sensitive to scales of the power
spectrum to which the COSEBIs are insensitive. We incorporate only a
single data point of the correlation function as this is sufficient to
capture the bulk of the additional information while simultaneously 
minimising possible B-mode contamination.

In order to determine the optimal scale of the data point that is to
be included, we consider 10 bins of $\xi_+$ ranging from 1.3 to 97.5
arcmin and perform 10 likelihood analyses for a combined data vector
consisting of 5 COSEBI modes and one additional data point of
$\xi_+$. We quantify the information content through the so-called
$q$ figure of merit ($q$-FoM)
\begin{eqnarray}
q &=& \sqrt{|\matQ|},
{\rm~~where}\nonumber \\
Q_{ij} &=& \int \mr d^2 \bpi \, p(\bpi| \,\vecd) \; (\pi_i-\pi_i^{\mr f})(\pi_j -\pi_j^{\mr f}) \,,
\label{eq:q_def}
\end{eqnarray}
$\bpi=(\Omega_m,\sigma_8)$ is the parameter vector, $p(\bpi|\vecd)$ is the posterior likelihood at this parameter point, and $\pi_i^{\mr f}$ denotes the fiducial parameter values. 
If the likelihood in parameter space (i.e. the posterior probability) is Gaussian, the $q$-FoM corresponds to the more common Fisher matrix based figure of merit $f=1/|\sqrt{\mathbf F}|$. The Fisher matrix $\mathbf F$ can be interpreted as the expectation value of the inverse parameter covariance evaluated at the maximum likelihood estimate parameter set, which in our ansatz corresponds to the fiducial parameters. Mathematically we can express this equivalence as
\begin{equation}
\label{eq:qf_equi}
f=\frac{1}{\sqrt{|\mathbf F|}}=\sqrt{|\matC_{\vpi}|}=\sqrt{|\matQ|}=q \,.
\end{equation}
Since the assumption of a Gaussian posterior is clearly violated in the $\sigma_8$-$\Omega_{\rm m}$ parameter space, we perform a full likelihood analysis and calculate $q$ to quantify the size of the likelihood. Note that smaller $q$-FoM is ``better.''

\begin{figure}
\includegraphics[width=7cm]{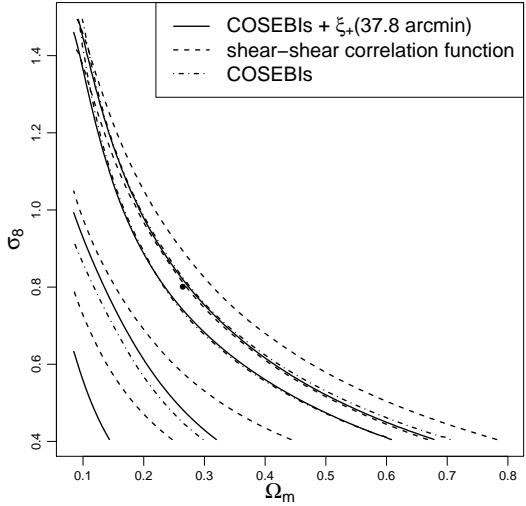}
\caption{\label{fig:likepanels}
The likelihood contours of the combined data vector (solid), the shear-shear correlation function (dashed), and the COSEBIs (dotted) data vector to illustrate how much information is gained when including the additional data point. Note that the COSEBIs' lower 95 per cent contour is outside the considered region. The dot indicates the WMAP7 best-fitting values.}
\end{figure}

We varied the angular scale (in arcmin) of the added $\xi_+(\theta)$ data point, and found a minimal $q$-FoM at $\theta=37.8$ arcmin. We will use this scale for the additional $\xi_+$ data point henceforth. Note that this analysis uses a simulated input data vector in order to avoid biases from designing a statistical test based on the observed data. The constraints coming from the various possible data vectors -- the COSEBIs, the COSEBIs supplemented with a single $\xi_+$ point, and the full shear correlation function -- are compared in Fig.~\ref{fig:likepanels}. They are not identical, which is expected since they weight the data in different ways, but are consistent with each other.


The COSEBI modes are highly correlated with each other, and they are
correlated to a lesser extent with $\xi_+$ at 38 arcmin. The
correlation matrix is shown in Fig.~\ref{fig:sixbysix}, and the
corresponding covariance matrix is tabulated in the Appendix in Table~\ref{table:covar}.

\begin{figure}
\includegraphics[angle=-90,width=3.2in]{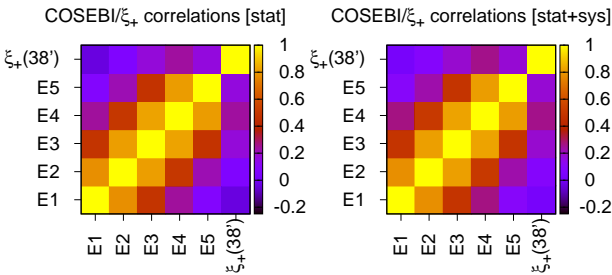}
\caption{\label{fig:sixbysix}The correlation matrix of the COSEBI modes 1--5 (``E1...E5'' in the figure) and $\xi_+(38')$. The left panel shows only the statistical (Monte Carlo) errors, and the right panel includes the systematics as well.}
\end{figure}

\subsection{Parameter Fits}
\label{ss:pfit}

\begin{figure}
\includegraphics[width=7cm]{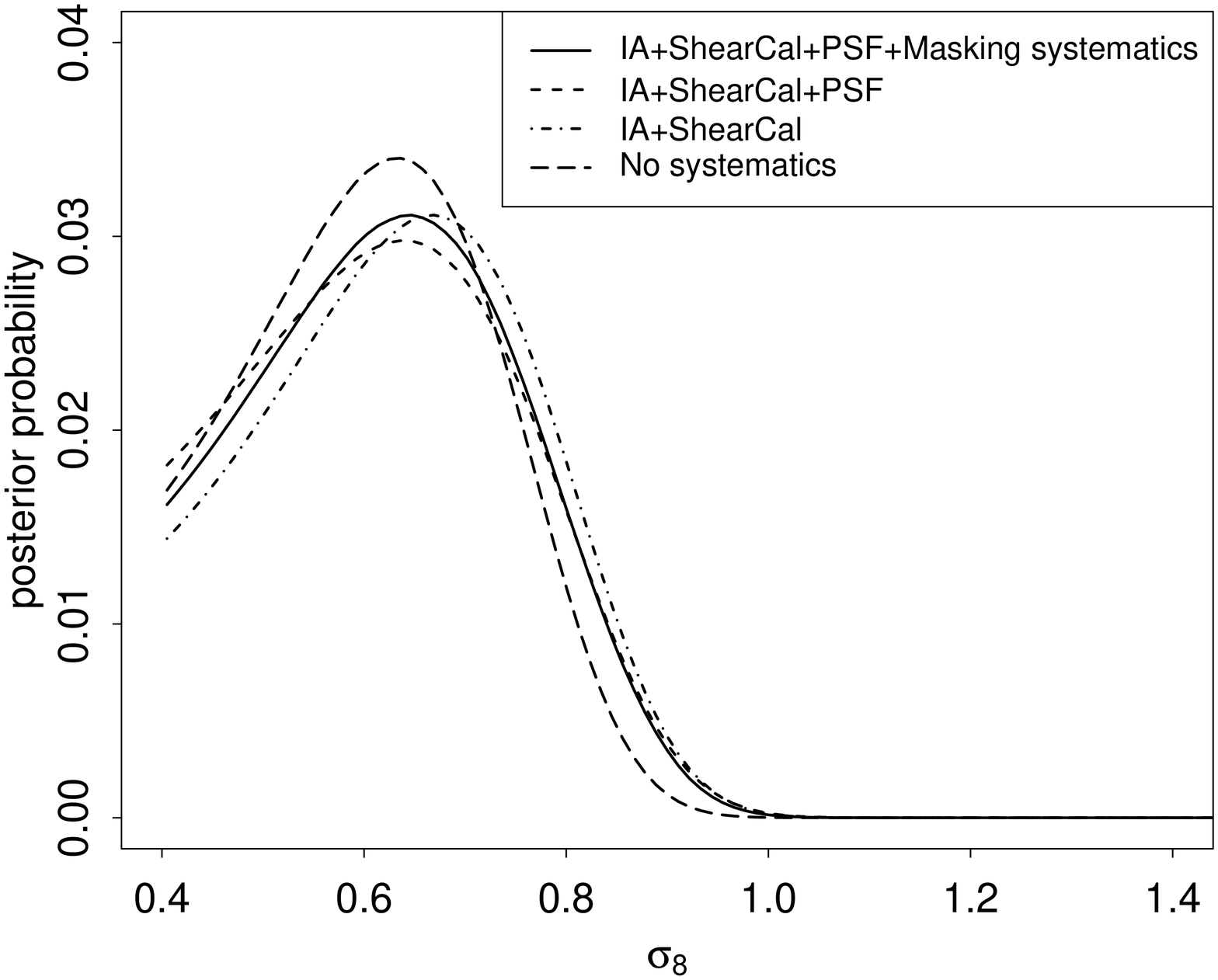}
\includegraphics[width=7cm]{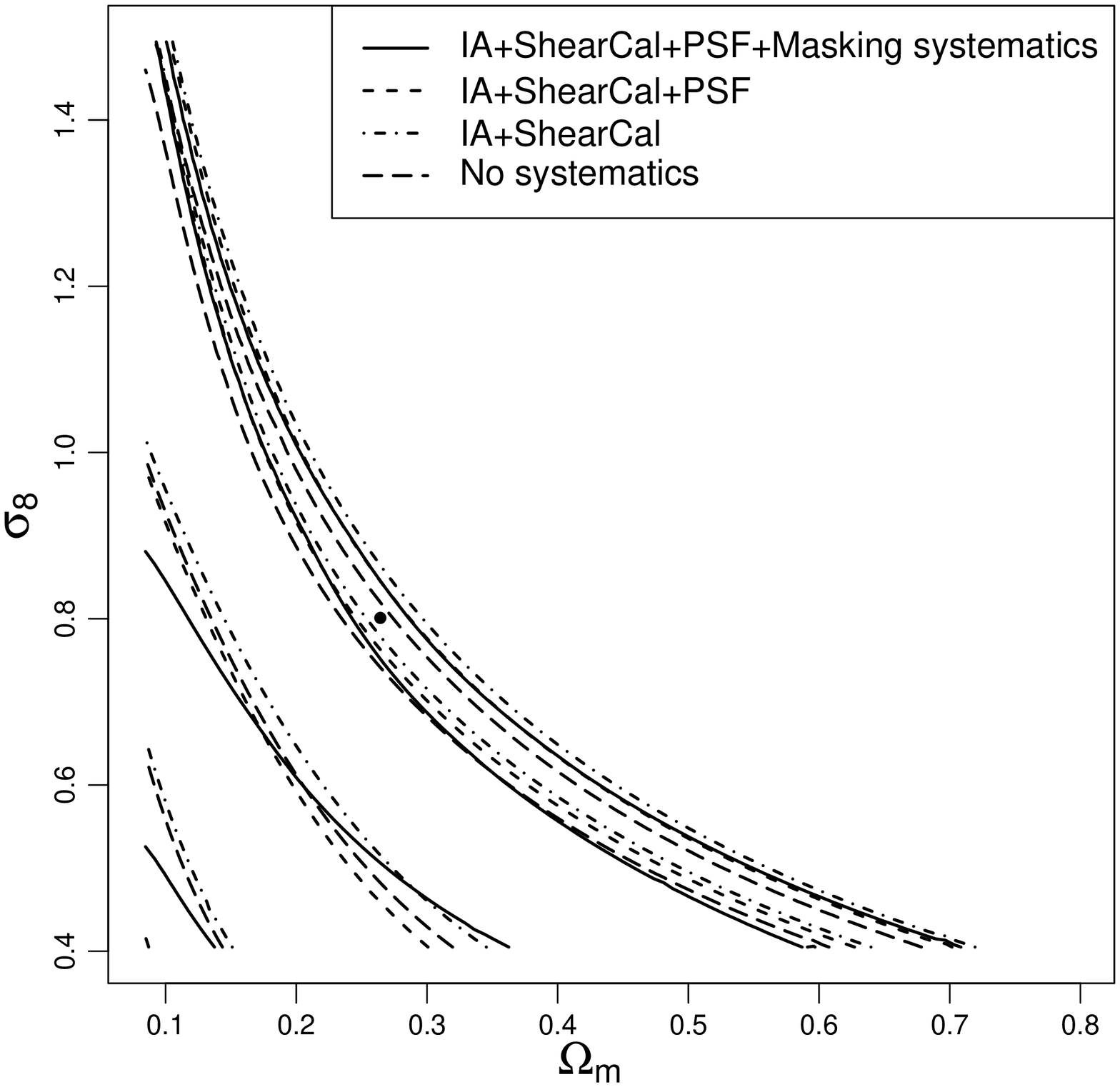}
\caption{\label{fig:syseffect}The effect of systematic errors in the
  1-D likelihood of $\sigma_8$ (upper panel) and in the 2-D
  constraints (68 per cent  likelihood contours only) in the
  $\sigma_8-\Omega_m$ plane (lower panel). The solid curve shows our
  final analysis, while the other curves show results including
  subsets of the systematic errors. The dot-dashed curve labeled ``no
  systematics'' shows only the statistical errors, without any
  systematic error corrections either to the theory or to the
  covariance matrix. The dot indicates the WMAP7 best-fitting values.}
\end{figure}

We perform all of our fits to a standard five-parameter $\Lambda$CDM
model\footnote{The optical depth to reionization $\tau$ is a sixth
  parameter implicitly included in the WMAP7 chains, but with no
  effect on the lensing shear correlation function.}. For the initial
likelihood analysis, we fix $n_s$, $\Omega_bh^2$, $\Omega_mh^2$, and
$w_0$ at their fiducial best-fit WMAP7 values
\citep{2011ApJS..192...18K}, and vary $\sigma_8$.  The upper panel of
Fig.~\ref{fig:syseffect} shows the likelihood of $\sigma_8$ with all
other parameters fixed, with a value at the peak and 68 per cent
confidence interval of $0.636^{+0.109}_{-0.154}$. For a survey of this
size and depth, the constraints are comparable to the statistically
achievable confidence limits.

We also perform a likelihood analysis fixing three parameters, and
varying $\Omega_m$ and $\sigma_8$ simultaneously, as these two
parameters are much more sensitive to the measured cosmic shear signal
than the others.  The resulting two-dimensional constraints are shown
in the bottom panel of Fig.~\ref{fig:syseffect}.  Our 68 per cent
confidence limits on the degenerate product
$\sigma_8\left(\frac{\Omega_m}{0.264}\right)^{0.67}$ are
$0.65^{+0.12}_{-0.15}$ for the Coyote Universe prediction code (see
Fig.~\ref{fig:syseffect}, solid red line), and
$\sigma_8\left(\frac{\Omega_m}{0.264}\right)^{0.72}=0.67^{+0.12}_{-0.15}$
for the Halofit prediction code (see Fig.~\ref{fig:syseffect}, dashed
red line).

We show the effects of removing each systematic error correction,
Fig.~\ref{fig:syseffect} also shows, for both the one- and
two-dimensional analyses, the impact of systematic error corrections. The combined effects of these uncertainties
are clearly substantially smaller than the statistical error on the
amplitude of the shear signal.

Finally, we adopt the WMAP7 likelihoods as priors, and evaluate our
likelihood at each link in the WMAP7 Markov chain. For each chain
element, we assign a weight equal to our likelihood function evaluated
at the parameter vector for that chain element. For each of the
parameter constraint plots shown here, we first assign each Markov
Chain Monte Carlo (MCMC) chain element to a point on a regular grid in
the parameter space; the value of the marginalised likelihood at each
grid-point, $H_{i,j}$ is then the sum of our likelihood weights over the MCMC
chain elements at the $(i,j)$ grid-point,
\begin{equation}
H_{i,j} = \sum_{k} I_k(i,j) L_k,
\end{equation}
where the indicator function $I_k(i,j)$ is equal to unity when the
$(i,j)$ grid-point in parameter space is nearest the $k$th chain
element, and zero otherwise. The likelihood $L_k$ for each chain element is evaluated in the usual way as:
\begin{equation}
L_k = \exp\left(-\frac{{\bar {\bmath d}_k}^T {\mathbfss C}^{-1} {\bar {\bmath d}_k} }{2}\right).
\end{equation}
Here ${\mathbfss C}$ is the full covariance matrix for the measurement,
incorporating both the statistical and systematic uncertainties, and
the normalization is arbitrary. The data vector ${\bar{\bmath d}}_k$ is the
extended COSEBI vector described above; where shown, the WMAP7 priors
are simply this sum with $L_k=1$ for each point.

We estimate the detection significance for the final signal, the
difference $\sqrt{-2\Delta \log L}$ between the highest-likelihood
Markov Chain element for both the $\Lambda$CDM and $w$CDM models and
the likelihood evaluated with no signal. The $1\sigma$ detection
significances for these two models are 2.64\ and 2.88,
respectively. This is not the significance of the detection of cosmic
shear (as in Sec.~\ref{subsec:syscontrib}), but rather a measurement of the likelihood of these two models
given the combination of WMAP7 priors with this experiment.

\begin{figure*}
\includegraphics[scale=0.4]{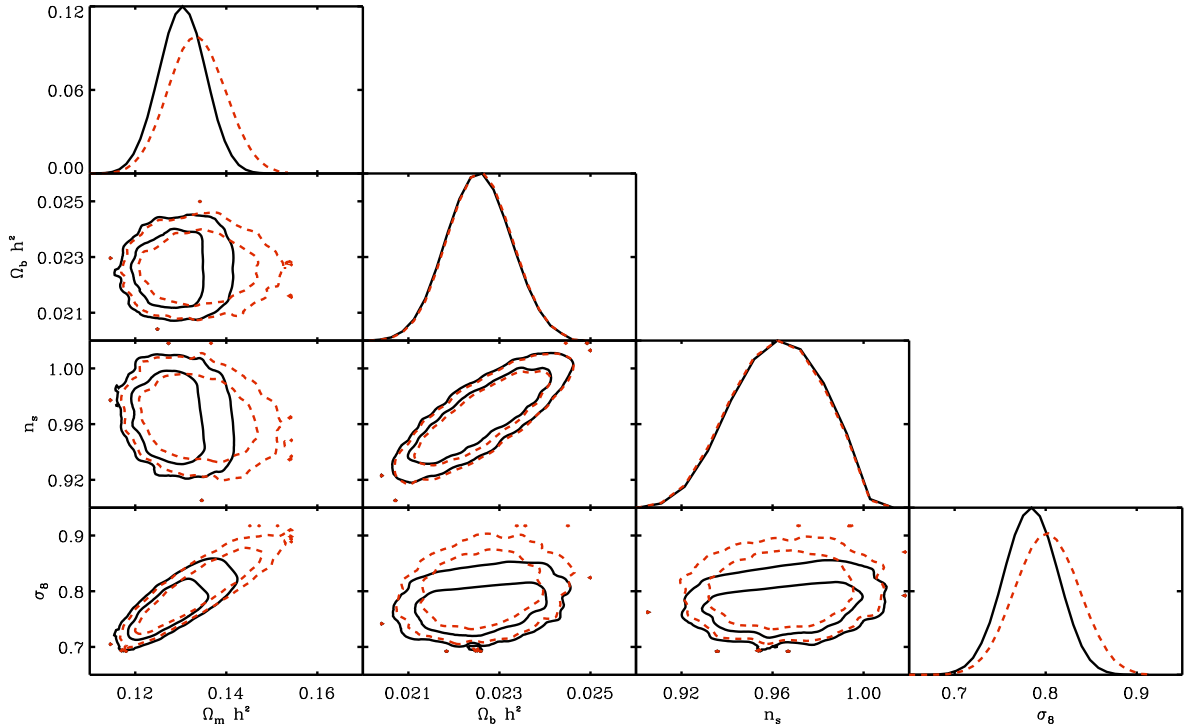}
\caption{\label{fig:LCDMconstraints} The cosmological parameter
  constraints using the extended COSEBI data vector, fixing the dark
  energy equation of state $w$ at $-1$, but allowing all other
  parameters to vary.  Off-diagonal panels show joint two-dimensional
  constraints after marginalization over all the other parameters,
  which are shown. For these, the red contours show the WMAP7 priors
  containing 68.5 and 95.4 per cent of the posterior probability. The
  black contours are the same but for WMAP7+SDSS lensing. Diagonal
  panels show the fully-marginalized one-dimensional posterior
  distribution for each parameter; for these panels, the red (dashed) contours
  show the marginalized WMAP7 constraints.}
\end{figure*}

\begin{figure*}
\includegraphics[scale=0.4]{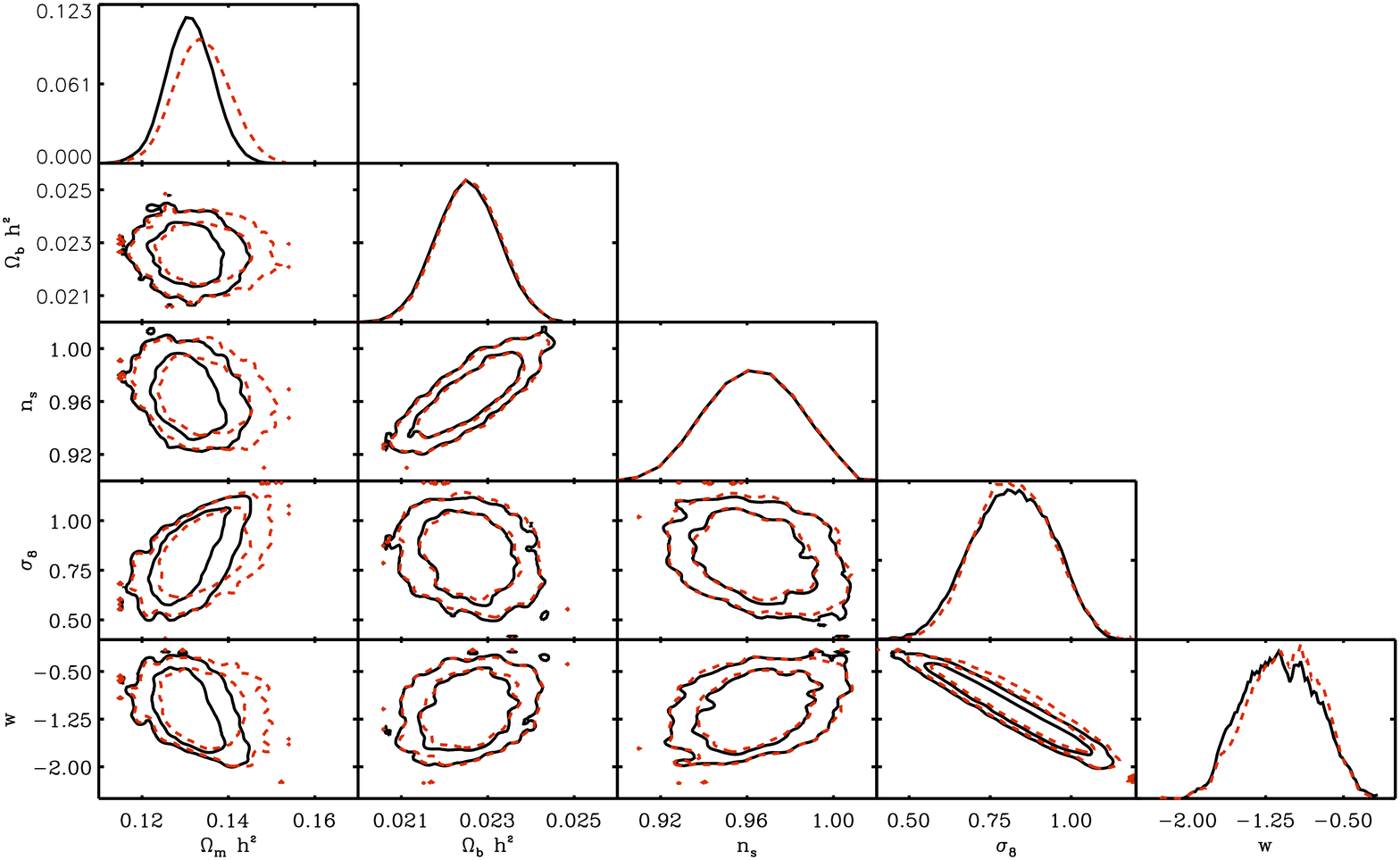}
\caption{\label{fig:wCDMconstraints} The cosmological parameter
  constraints using the extended COSEBI data vector, varying all five
  parameters.  Off-diagonal panels show joint two-dimensional
  constraints after marginalization over all the other parameters,
  which are shown. For these, the red contours show the WMAP7 priors
  containing 68.5 and 95.4 per cent of the posterior probability. The
  black contours are the same but for WMAP7+SDSS lensing. Diagonal
  panels show the fully-marginalized one-dimensional posterior
  distribution for each parameter; for these panels, the red (dashed) contours
  show the marginalized WMAP7 constraints.}
\end{figure*}

In Fig.~\ref{fig:LCDMconstraints}, we show marginalized posterior
likelihoods in the case of fixed $\Lambda$CDM (i.e., $w=-1$) for
$\Omega_m h^2$, $\Omega_b h^2$, $n_s$, and $\sigma_8$.  The results
with a free equation of state of dark energy (wCDM) are in
Fig.~\ref{fig:wCDMconstraints}.  Our measurement provides some
additional constraints beyond those from WMAP7 on these parameters. In
particular, the low amplitude of the measured shear signal rules out
some of the previously allowed volume of $\Omega_{\rm m}h^2$ and
$\sigma_8$ WMAP7 constraints.

\section{Conclusions} \label{sec:conclusions}

Using coadded imaging constructed from SDSS Stripe 82 data, we
constructed a weak lensing catalogue of 1\,328\,885 galaxies covering
168 square degrees (Paper I), and showed that the additive shear
systematics arising from the PSF are negligible compared to the cosmic
shear signal. In this paper, we carried out a cosmic shear measurement
that resulted in a 20\ per cent constraint on $\sigma_8$ (with all
other cosmological parameters fixed).  This adds constraining power
beyond that from WMAP7, and serves as an important independent data
point on the amplitude of the matter power spectrum at late times. In
particular, the primary CMB anisotropies presently provide only a
modest constraint on $\Omega_mh^2$, and (due to the effect of matter
density on the growth of structure) there is then an elongated allowed
region in the $(\Omega_mh^2,\sigma_8)$ plane; see
Fig.~\ref{fig:LCDMconstraints}. The WMAP7-allowed region is ideally
oriented for lensing to play a role: the lensing signal at the
high-$\Omega_mh^2$, high-$\sigma_8$ end of the ellipse leads to a much
higher lensing signal than low $\Omega_mh^2$, low $\sigma_8$. The low
amplitude of cosmic shear observed in this paper eliminates the
high-$\Omega_mh^2$, high-$\sigma_8$ solutions, and leads to a
WMAP7+SDSS lensing solution of
$\sigma_8=0.784^{+0.028}_{-0.026}$(1$\sigma$)$^{+0.055}_{-0.054}$(2$\sigma$)
and
$\Omega_mh^2=0.1303^{+0.0047}_{-0.0048}$(1$\sigma$)$^{+0.0091}_{-0.0092}$(2$\sigma$);
the 2$\sigma$ error ranges are respectively 14\ and 17\ per
cent smaller than for WMAP7 alone.

We have also carefully evaluated other sources of uncertainty such as
the source redshift distribution, intrinsic alignments, and shear
calibration, to ensure that our measurement is dominated by
statistical errors rather than systematic errors.  This achievement is
important when considering that (i) the SDSS data were never designed
with this application in mind, and indeed includes several features
(e.g. the minimal amount of cross-scan dithering) that cause significant
difficulty, and (ii) with the multitude of upcoming multi-exposure lensing surveys in
the next few years, it is important to cultivate new data analysis
techniques (such as the one used here) that are capable of producing
homogeneous data with tight control over PSF anisotropies. As a quantitative
measure of the extent of PSF correction possible with SDSS data, we take the
RMS residual spurious shear at a particular scale estimated from the star-galaxy
correlations,
\begin{equation}
\gamma_{\rm rms,eq}(\theta) = \frac{\sqrt{{\cal R}_{\rm psf}\,\xi_{+,\rm sg}(\theta)}}{{\cal R}}.
\end{equation}
From Fig.~\ref{fig:sgxf_w}, we see that this is $\sim 2\times 10^{-3}$
at the smallest scales (1--6 arcmin), is $<10^{-3}$ at scales
$\theta>0.1\,$degree, and drops to $3.7\times 10^{-4}$ in the final
bin (1.2--2.0 degrees).\footnote{We used ${\cal R}_{\rm psf}=0.9$ and
  ${\cal R}=1.776$, as described in the text.} There is almost no
difference between the $++$ and $\times\times$ signals, suggesting
that the spurious additive ellipticity signal contains similar amounts
of $E$- and $B$-modes\footnote{Recall that
  $\xi_{++}(\theta)-\xi_{\times\times}(\theta)$ and
  $P_E(\ell)-P_B(\ell)$ are $J_4$ Hankel transforms of each other.};
something similar was seen in the SDSS single-epoch data via
run-by-run comparisons of ellipticity measurements on the same
galaxies \citep[Fig. 8]{2006MNRAS.370.1008M}. This is good news for
the use of the $B$-mode as a diagnostic of PSF systematics, although
an understanding of the generality of this pattern remains elusive.

A major lesson learned from this project is the importance of {\em
  masking bias}, in which the intrinsic orientation of a galaxy
affects whether it falls within the survey mask. This is likely the
main reason why we had to implement the $\langle e_1\rangle$
projection. While we have clearly not exhausted the range of options
for removing this bias at the catalogue level, future surveys should
be designed to produce more uniform data quality via an appropriate
dithering strategy and suppress the masking bias at the earliest
stages of the analysis.

Our major limitation in the end was the source number density, which
was driven by the fact that our PSF-matching procedure was limited by
the worst seeing in the images that we use, and therefore we had to
eliminate the images with seeing worse than the median.  This means
that the coadds were not as deep as they could have been, and the
final effective seeing was 1.31 arcsec (full-width half maximum).  In principle this will be an
obstacle to applying this technique in the future, but in fact, that
statement depends on context.  For example, for a survey such as HSC or LSST
where we expect typically $\sim 0.7$ arcsec seeing, and with plans to
preferentially use the best-seeing nights for $r$ and $i$-band imaging that
will be used for shape measurement, it is conceivable that nearly all
images intended for lensing will have seeing in the 0.6--0.8
arcsec range.  In that context, a PSF-matched coadd that has the rounding
kernel applied may actually not result in much loss of information
about the shapes of most useful galaxies, and will have the advantage
of the removal of PSF anisotropies.  Moreover, even for surveys for
which the loss of information that results from this method may not be
suitable for the final cosmological analysis, this method may still
serve as a useful diagnostic of the additive PSF systematics.

\section*{Acknowledgments}

We thank Alexie Leauthaud for providing faint COSMOS galaxy postage
stamp images for simulation purposes.  E.M.H. is supported by the US Department of Energy's Office of High Energy Physics (DE-AC02-05CH11231).
During the period of work on this paper, C.H. was supported by the US
Department of Energy's Office of High Energy Physics (DE-FG03-02-ER40701 and DE-SC0006624), the US National Science
Foundation (AST-0807337), the Alfred P. Sloan Foundation, and the
David \& Lucile Packard Foundation. R.M. was supported for part of
the duration of this project by NASA
through Hubble Fellowship grant \#HST-HF-01199.02-A awarded by the
Space Telescope Science Institute, which is operated by the
Association of Universities for Research in Astronomy, Inc., for NASA,
under contract NAS 5-26555.  U.S. is supported by the DOE, the Swiss National Foundation under contract 200021-116696/1 and WCU grant R32-10130.

We thank the PRIMUS team for sharing their redshift catalogue, and thank Alison Coil and
John Moustakas for help with using the PRIMUS dataset.  Funding for
PRIMUS has been provided by NSF grants AST-0607701, 0908246, 0908442,
0908354, and NASA grant 08-ADP08-0019.  This paper includes data
gathered with the 6.5 meter Magellan Telescopes located at Las
Campanas Observatory, Chile.
     
Funding for the DEEP2 survey has been provided by NSF grants
AST95-09298, AST-0071048, AST-0071198, AST-0507428, and AST-0507483 as
well as NASA LTSA grant NNG04GC89G.  Some of the data presented herein
were obtained at the W. M. Keck Observatory, which is operated as a
scientific partnership among the California Institute of Technology,
the University of California and the National Aeronautics and Space
Administration. The Observatory was made possible by the generous
financial support of the W. M. Keck Foundation. The DEEP2 team and
Keck Observatory acknowledge the very significant cultural role and
reverence that the summit of Mauna Kea has always had within the
indigenous Hawaiian community and appreciate the opportunity to
conduct observations from this mountain. 

Funding for the SDSS and SDSS-II has been provided by the Alfred
P. Sloan Foundation, the Participating Institutions, the National
Science Foundation, the U.S. Department of Energy, the National
Aeronautics and Space Administration, the Japanese Monbukagakusho, the
Max Planck Society, and the Higher Education Funding Council for
England. The SDSS Web Site is http://www.sdss.org/. 

The SDSS is managed by the Astrophysical Research Consortium for the
Participating Institutions. The Participating Institutions are the
American Museum of Natural History, Astrophysical Institute Potsdam,
University of Basel, University of Cambridge, Case Western Reserve
University, University of Chicago, Drexel University, Fermilab, the
Institute for Advanced Study, the Japan Participation Group, Johns
Hopkins University, the Joint Institute for Nuclear Astrophysics, the
Kavli Institute for Particle Astrophysics and Cosmology, the Korean
Scientist Group, the Chinese Academy of Sciences (LAMOST), Los Alamos
National Laboratory, the Max-Planck-Institute for Astronomy (MPIA),
the Max-Planck-Institute for Astrophysics (MPA), New Mexico State
University, Ohio State University, University of Pittsburgh,
University of Portsmouth, Princeton University, the United States
Naval Observatory, and the University of Washington.

\appendix
\section{The Data Vector and Covariance Matrix.}
Here we reprint the data vector and covariance matrix used in this
measurement. The code used to project the correlation function onto
the COSEBI basis functions is available from the authors upon request.

\begin{table}
  \caption{Our data vector. The first five elements are COSEBI mode amplitudes; the final is the correlation function averaged in the range $29.2296 \le \vt \le 44.9730 $.}
\begin{tabular}{l}
  \hline\hline
4.89797E-10 \\
1.28335E-09 \\
1.25136E-09 \\
1.45616E-09 \\
8.92333E-10 \\
1.46457E-05 \\
  \hline\hline
\label{table:datavector}
\end{tabular}
\end{table}

\begin{table}
  \caption{The covariance matrix for the data vector shown in table \ref{table:datavector}.}
\begin{tabular}{c c c}
  \hline\hline
  Data vector index & Data vector index & Covariance \\
  \hline\hline
       0 &   0 &  3.37161E-20 \\
       0 &   1 &  4.67637E-20 \\
       0 &   2 &  4.00484E-20 \\
       0 &   3 &  2.49916E-20 \\
       0 &   4 &  9.84257E-21 \\
       0 &   5 &  3.01770E-17 \\
       1 &   1 &  1.06383E-19 \\
       1 &   2 &  1.19226E-19 \\
       1 &   3 &  8.39508E-20 \\
       1 &   4 &  3.86519E-20 \\
       1 &   5 &  1.82344E-16 \\
       2 &   2 &  1.99923E-19 \\
       2 &   3 &  1.87469E-19 \\
       2 &   4 &  1.12196E-19 \\
       2 &   5 &  5.07790E-16 \\
       3 &   3 &  2.56568E-19 \\
       3 &   4 &  2.13363E-19 \\
       3 &   5 &  8.02118E-16 \\
       4 &   4 &  2.67774E-19 \\
       4 &   5 &  5.67797E-16 \\
       5 &   5 &  3.68112E-11 \\
  \hline\hline
\label{table:covar}
\end{tabular}
\end{table}

\end{document}